\documentclass[
  5p,
  final,
  times,
  sort&compress,
]{elsarticle}

\journal{Nucl.~Instrum.~Methods Phys.~Res., Sect. A}

\usepackage{setspace}
\usepackage{amsmath}
\usepackage{amssymb}
\usepackage{booktabs}
\usepackage{tabularx}
\usepackage{dcolumn}
\usepackage{caption}
\usepackage{textgreek}
\usepackage[T1]{fontenc}
\usepackage[utf8]{inputenc}
\usepackage{textcomp}
\usepackage{mathtools}

\usepackage{microtype}
\microtypesetup{%
  protrusion=true,
  expansion=true,
  tracking=true,
  kerning=true,
}
\microtypecontext{spacing=nonfrench}

\usepackage{hyperref}
\hypersetup{
  bookmarksopen=true,
  bookmarksopenlevel=2,
  bookmarksnumbered=true,
  colorlinks=true,
  pdfdisplaydoctitle=true,
  pdffitwindow=true,
}
\usepackage{siunitx}
\newcommand{\MeV}{\mega\electronvolt}
\DeclareSIUnit\Tm{\mathrm{Tm}}
\DeclareSIUnit\mrad{\milli\radian}
\DeclareSIUnit\msr{\milli\steradian}
\DeclareSIUnit\gauss{G}

\makeatletter
\g@addto@macro{\UrlBreaks}{%
\do\/%
\do\a\do\b\do\c\do\d\do\e\do\f\do\g\do\h\do\i\do\j\do\k\do\l\do\m%
\do\n\do\o\do\p\do\q\do\r\do\s\do\t\do\u\do\v\do\w\do\x\do\y\do\z%
\do\A\do\B\do\C\do\D\do\E\do\F\do\G\do\H\do\I\do\J\do\K\do\L\do\M%
\do\N\do\O\do\P\do\Q\do\R\do\S\do\T\do\U\do\V\do\W\do\X\do\Y\do\Z%
}
\makeatother

\usepackage{etoolbox}
\makeatletter
\patchcmd\ps@pprintTitle
  {\fi\fi\fi\fi}
  {\fi\fi\fi\fi\afterassignment\fix@elsarticle\let\@tempa}
  {}{\FailedToPatch}
\def\fix@elsarticle{\iffalse{\fi}\romannumeral-`0}
\makeatother

\begin{document}

\begin{frontmatter}

\title{Design of the High Rigidity Spectrometer at FRIB}

\author[FRIB]{\texorpdfstring{S.~Noji\corref{correspondingauthor}}{}}
\ead{noji@frib.msu.edu}
\cortext[correspondingauthor]{Corresponding author}
\author[FRIB,PA]{R.~G.~T.~Zegers}
\author[ND]{G.~P.~A.~Berg}
\author[BU]{A.~M.~Amthor}
\author[FRIB]{T.~Baumann}
\author[FRIB,PA]{D.~Bazin}
\author[FRIB,ORNL]{E.~E.~Burkhardt}
\author[FRIB]{M.~Cortesi}
\author[FRIB]{J.~C.~DeKamp}
\author[FRIB]{M.~Hausmann}
\author[FRIB]{M.~Portillo}
\author[ANL]{D.~H.~Potterveld}
\author[FRIB,PA]{B.~M.~Sherrill}
\author[FRIB]{A.~Stolz}
\author[FRIB]{O.~B.~Tarasov}
\author[FRIB]{R.~C.~York}

\address[FRIB]{Facility for Rare Isotope Beams, Michigan State University, East Lansing, MI 48824 USA}
\address[PA]{Department of Physics and Astronomy, Michigan State University, East Lansing, MI 48824 USA}
\address[ND]{Department of Physics, University of Notre Dame, Nieuwland Science Hall, Notre Dame, IN 46556 USA}
\address[BU]{Department of Physics and Astronomy, Bucknell University, Lewisburg, PA 17837 USA}
\address[ORNL]{Fusion Energy Division, Oak Ridge National Laboratory, Oak Ridge, TN 37831 USA}
\address[ANL]{Physics Division, Argonne National Laboratory, Argonne, IL 60439 USA}

\begin{abstract}
A High Rigidity Spectrometer (HRS) has been designed for experiments at the Facility for Rare-Isotope Beams (FRIB) at Michigan State University (MSU).
The HRS will allow experiments to be performed with the most exotic neutron-rich isotopes at high beam energies (\SI{\gtrsim100}{\MeV}/$u$).
The HRS consists of an analysis beamline called the High-Transmission Beamline (HTBL) and the spectrometer proper called the Spectrometer Section.
The maximum magnetic rigidity of the HRS is \SI{8}{\Tm}, which corresponds to the rigidities at which rare-isotope beams are optimally produced at FRIB.
The resolving power, angular acceptance, and momentum acceptance are set to match the anticipated scientific program.
An ion-optical design developed for the HRS is described in detail, along with the specifications of the associated magnet and detector systems.
\end{abstract}


\end{frontmatter}

\section{Introduction}

\begin{figure*}[t]
  \centering
  \includegraphics{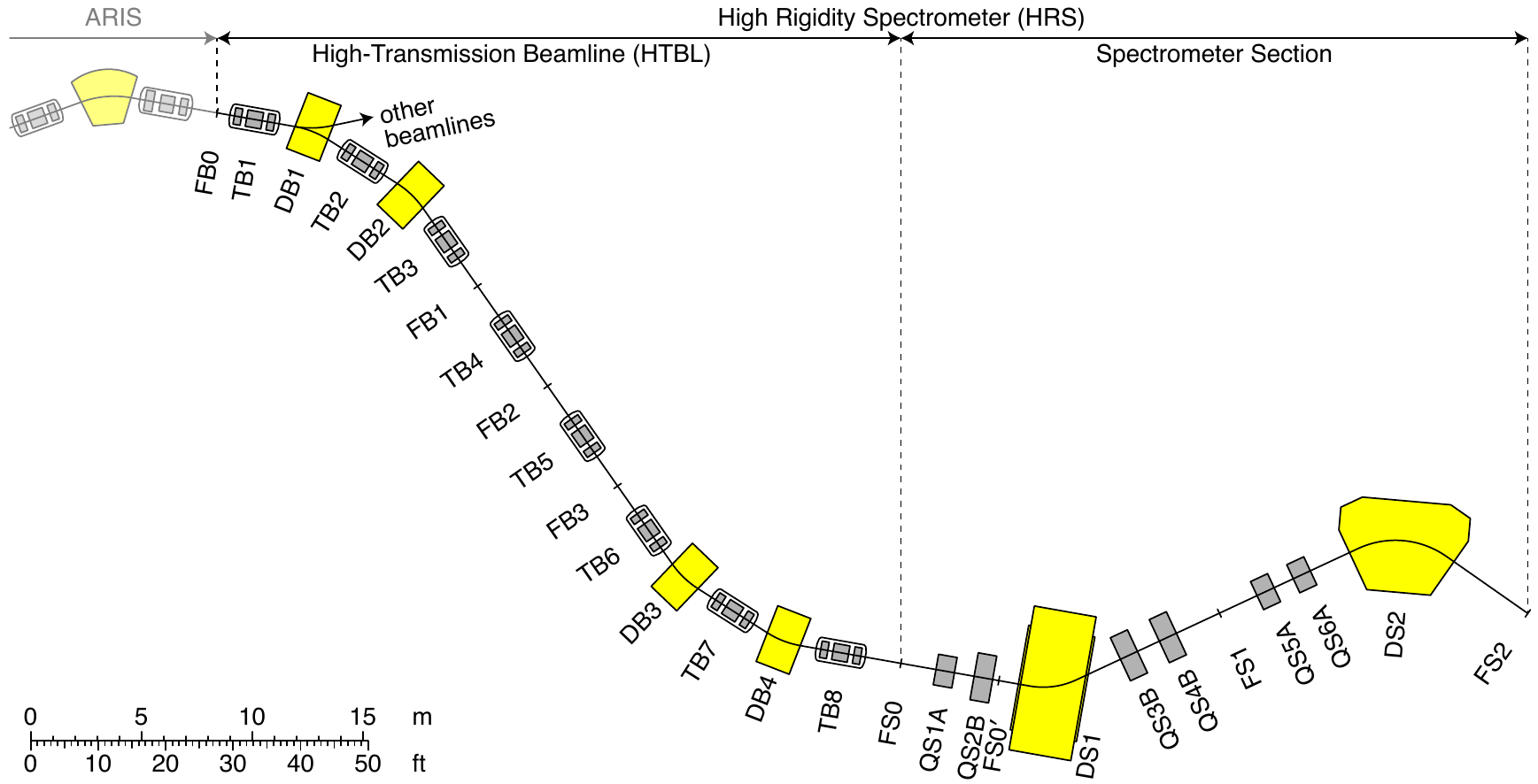}
  \caption{%
    Layout of the HRS consisting of the HTBL and the Spectrometer Section.
    The yellow and gray boxes represent dipole and quadrupole magnets, respectively.
    The naming convention is as follows:  the first letter denotes the element type, i.e.~``F'' for a focal plane, ``D'' for a dipole magnet, ``Q'' for a quadrupole magnet, ``T'' for a quadrupole triplet; the second letter ``B'' and ``S'' denotes the HTBL and the Spectrometer Section, respectively; the following sequential numbers start at 0 for the focal planes, and 1 for the other elements; the quadrupoles of the Spectrometer Section have additional type identifiers (``A'' and ``B'').
    The HTBL starts after the final image of ARIS (FB0), and the other beamlines branch off from DB1.
  }
  \label{fig:layout}
\end{figure*}

The Facility for Rare Isotope Beams (FRIB)~\cite{JInst.15.P12034,ModPhysLettA.2230006} at Michigan State University (MSU) is one of the world's premier rare-isotope-beam experimental facilities, capable of producing about \SI{80}{\%} of the isotopes of elements up to uranium that are predicted to exist~\cite{Nature.486.509,PhysLettB.726.680}.
Unprecedented access to yet-to-be-discovered isotopes will deepen our understanding of the fundamental forces that bind nucleons within nuclei, the astrophysical origin of nuclei, and the properties and interactions of rare isotopes.

A new system of a magnetic spectrometer and an associated analysis beamline, called the High Rigidity Spectrometer (HRS), which accommodates magnetic rigidities up to \SI{8}{\Tm}, has been designed for experiments with fast (\SI{\gtrsim100}{\MeV}/$u$) beams at FRIB.
Such a system will substantially enhance the scientific reach of FRIB's experimental program, compared to that by only using the existing devices at the National Superconducting Cyclotron Laboratory (NSCL) at MSU, namely the S800 Spectrograph~\cite{NIMB.204.629} and the Sweeper Magnet~\cite{IEEETransApplSupercon.15.1252}, which have the maximum magnetic rigidities of \SI{4}{\Tm}.

The higher magnetic rigidity of the HRS will match the scientific program enabled by the \SI{200}{\MeV}/$u$ superconducting heavy-ion driver linear accelerator and the Advanced Rare-Isotope Separator (ARIS), whose maximum magnetic rigidity is \SI{8}{\Tm}~\cite{NIMB.317.349,NIMB.376.150}.
The HRS will allow experiments to run with beams at magnetic rigidities at which their production rates via projectile fragmentation or in-flight fission are optimized.
The design magnetic rigidity of the HRS is likewise suitable for the envisioned FRIB energy upgrade from \SI{200}{MeV}/$u$ to \SI{400}{MeV}/$u$~\cite{JInst.15.P12034}.

For rare-isotope experiments, which are often performed with very weak beams, luminosity is a critical parameter.
Two factors contribute to luminosity, namely the rare-isotope beam intensity at the reaction target and the thickness of the reaction target.
The beam intensity at the target is determined by the beam intensity produced in ARIS and the beam transmission efficiency in the subsequent beamline to the target.
To optimize the latter, high magnetic rigidity is crucial as it enables beams to be transported at their optimal energies without having to degrade them.
This is important because energy degradation will result in increased emittances and hence reduced intensities.
Use of thicker reaction targets is also enabled by higher beam energies owing to smaller fractional energy losses.
The combined gain in luminosity from these two factors will be up to two orders of magnitude over the existing devices, and the largest gains are for the most neutron-rich rare-isotope beams~\cite{HRSPreliminaryDesignReport}.

The higher magnetic rigidity allowing experiments to run at the beam-energy range of \SIrange[range-phrase={--},range-units=single]{100}{250}{\MeV}/$u$ is also optimal for studies of various types of direct reactions, such as knockout or charge-exchange reactions, and enables new reactions such as quasi-free proton scattering.
Performing experiments in this energy range will improve the reliability of reaction theory because necessary approximations used in interpretation of those experiments become more appropriate.

The HRS is designed as a versatile spectrometer system for a wide variety of experiments.
It consists of a spectrometer proper called the Spectrometer Section and its preceding analysis beamline called the High-Transmission Beamline (HTBL), which starts after the final image of ARIS.
The layout is shown in Fig.~\ref{fig:layout}.
While earlier conceptual ion-optical designs have been presented elsewhere~\cite{NIMB.376.162,IJMPA.34.1942017,osti_1573440,HRSPreliminaryDesignReport}, a more advanced design is described in this paper.
The design objectives driven by the science program are discussed in Section~\ref{section:specifications}.
Details of the ion-optical designs of the HTBL and the Spectrometer Section that achieve these objectives are described in Sections~\ref{section:ionOpticalDesignHTBL} and \ref{section:ionOpticalDesignSpectrometerSection}, respectively.
The specifications of the magnets and the detectors necessary to realize the ion-optical design are stipulated in Sections~\ref{section:magnets} and \ref{section:detectors}, respectively.

Most of the ion-optical calculations presented in this paper have been performed using COSY Infinity (simply referred to as COSY hereafter)~\cite{NIMA.558.346,COSY}, which can compute transfer maps for arbitrarily complicated fields up to arbitrary order.
Those COSY calculations included realistic three-dimensional field profiles including the fringe fields of the magnets.
A toolkit developed for the ion-optical development of ARIS, as described in Ref.~\cite{NIMB.376.150}, was used to enable Monte-Carlo simulations of beam transport in LISE$^{++}_{\mathit{cute}}$~\cite{LISECUTE,NIMA.482.307,NuclPhysA.746.411,NIMB.266.4657,NIMB.376.185,JPhysConfSer.664.072029,NIMB.376.168} using COSY-calculated transfer maps.
The indices of the maps calculated in COSY~\cite{Berz.AIEP108book,COSYBeamMan91} are the positions in the dispersive ($x$) and non-dispersive ($y$) planes, and the normalized momenta $a \equiv p_x/p_0$ and $b \equiv p_y/p_0$, which are the ratios of the dispersive ($p_x$) and non-dispersive ($p_y$) momentum to the total momentum ($p_0$). 
Here, $x$--$a$ and $y$--$b$ are canonically conjugate to each other, and $a$ and $b$ are approximately equal to the dispersive and non-dispersive geometrical angles ($x^\prime$ and $y^\prime$) when they are small. 
While COSY uses $d \equiv (K-K_0)/K_0$ to denote the deviation of the kinetic energy $K$ from the central kinetic energy $K_0$, the corresponding momentum deviation $\varDelta p/p_0 \equiv (p-p_0)/p_0$ is used in this paper unless otherwise noted.

\section{Design objectives}
\label{section:specifications}

The HRS will enable a wide variety of experiments at FRIB \cite{AnnRevNuclPartSci.56.53,osti_1296778}.
The objectives of these experiments define the goals for the HRS specifications.
These specifications can be categorized into two sets, namely one that pertains to the Spectrometer Section, as described in Section~\ref{subsection:specificationsSpectrometerSection}, and the other that pertains to the HTBL, as described in Section~\ref{subsection:specificationsHTBL}.

\subsection{Spectrometer Section}
\label{subsection:specificationsSpectrometerSection}

The Spectrometer Section has different modes of operation to accommodate various experimental needs.
The principal mode, which is referred to as the high-resolution mode, will be used for the majority of the experiments.
These experiments are a natural extension of what has been performed at the S800.
They also include mass-measurement experiments where nuclear masses are determined by measuring the magnetic rigidity ($B\rho$) and the time-of-flight (ToF) over a long flight path.
For those mass measurements the entire HRS will be used as a single achromatic spectrometer, in the same way as what has been done with the S800.
The other mode is called the neutron-invariant-mass mode, which is specialized for invariant-mass spectroscopy involving detection of fast neutrons, which has been performed with the Sweeper Magnet.

In the following, the necessary performance is discussed in the framework of these operational modes.
Because many experiments have overlapping performance specifications, several prototypical experiments that have the most stringent specifications encompassing those of the other experiments are used to define those specifications.

\subsubsection{High-resolution mode}

Knockout reactions from the heaviest nuclei of mass number of up to \num{238} set a constraint that the momentum resolving power must be better than $1/\num{1500} = 0.067\%$ (FWHM) so that the angular momentum carried by the knocked-out nucleon can be determined.
Particle identification of the reaction residue in the spectrometer is done by measuring its ToF.
Assuming that the resolution of the timing detector is \SI{150}{\ps} (FWHM), a flight path of at least \SI{25}{\meter} is necessary to achieve a $4 \sigma$ mass resolution at the beam energy of about \SI{200}{\MeV}/$u$.
Based upon the experience with the S800 experiments, to profile the momentum distribution of the reaction residue efficiently in a single magnetic-rigidity setting of the spectrometer, a momentum acceptance of \SI{\pm2.5}{\%} is necessary.

Dispersion matching~\cite{Hendrie.NuclearSpectroscopyReactions,NIM.214.281,NIMB.126.274,NIMA.484.17,NIMB.266.4201}, where the entire HRS is set to be achromatic, is a powerful technique for achieving good momentum resolution.
Also, because dispersion matching causes the unreacted beam to be achromatically focused at the focal plane, it facilitates efficient blocking of the unreacted beam preventing it from reaching the focal-plane detectors.
Such a beam tune is particularly important for experiments with heavy beams, because unreacted beam particles are closer to reaction products of interest, even within the momentum acceptance of the spectrometer, which may, without such blocking, limit beam intensities with which experiments can run.
Moreover, equipping the spectrometer with an additional intermediate focal plane makes it possible to achieve dispersion matching to the intermediate focal plane before it reaches the final focal plane.
Such dispersion matching will be hereafter referred to as partial dispersion matching in contrast to full dispersion matching to the final focal plane.

In-beam {\textgamma}-ray spectroscopy using FRIB beams with the Gamma-Ray Energy Tracking Array (GRETA)~\cite{GRETAFinalDesignReport} is an important motivation for constructing the HRS.
Combining the HRS and GRETA will create the world's most powerful in-beam {\textgamma}-ray spectroscopy facility.
To realize this, the HRS must be able to accommodate at the target station (FS0 in Fig.~\ref{fig:layout}) GRETA with all the detector modules.
Because GRETA is the largest auxiliary detector system foreseen to be used around the target, a wide variety of other auxiliary detector systems can also be accommodated.

In-flight fission experiments where multiple fission products are to be measured simultaneously set a constraint on the solid-angle acceptance so that a sizable fraction of the fission products can be captured.
As stated above, a $4 \sigma$ mass resolution can be achieved for the mass number up to \num{238} at the beam energy of \SI{200}{\MeV}/$u$, with a flight path of \SI{25}{\meter} or longer and the resolution of the timing detector of \SI{150}{\ps} (FWHM).
To achieve $4 \sigma$ separation between isotopes with neighboring charge number up to $Z = 92$, a resolution in energy loss of \SI{1.3}{\%} (FWHM) is needed for an energy-loss detector.
To evaluate the detection efficiency, simulations were performed in LISE$^{++}_{\mathit{cute}}$ with a \SI{200}{\MeV}/$u$ beam of ${}^{238}\mathrm{U}$, assuming an excitation energy of \SI{20}{\MeV}.
A momentum acceptance was set at \SI{2.5}{\%}, and the solid-angle acceptance at \SI{15}{\msr}.
About \SI{10}{\%} efficiency was obtained for the simultaneous detection of two fission products, averaged over mass numbers, which is sufficient to enable efficient in-flight fission studies.
Increasing of solid-angle coverage further would not improve the detection efficiency since the momentum acceptance would then become the constraining factor~\cite{HRSPreliminaryDesignReport}.

ToF-$B\rho$ mass measurements utilize the long flight path of the HRS (\SI{\sim70}{\meter} long), which is rendered an achromatic spectrometer as a whole by the dispersion-matching technique.
As in the experiments performed at the S800~\cite{PhysRevLett.107.172503,NIMA.696.171,IJMS.349.145,PhysRevLett.114.022501}, where the beamline preceding the S800 was added to make a \SI{58.7}{\meter} flight path, the second half of ARIS (the reconfigured A1900 Fragment Separator after the Preseparator) will be added to the HRS to make a \SI{\sim100}{\meter} flight path.
Dedicated timing detectors with a resolution of \SI{30}{\ps} (FWHM)~\cite{NIMA.974.164199,NIMA.1027.166050} are foreseen to be used to achieve a mass resolving power of $\varDelta m / m \sim \num[retain-unity-mantissa = false]{1e-4}$ in the FRIB energy region.

\subsubsection{Neutron-invariant-mass mode}

Neutron-invariant-mass experiments involve measurements of momentum vectors of charged fragments and of neutrons resulting from the breakup of unbound states and emitted at forward angles.
These neutrons will be detected by neutron detectors such as the MoNA-LISA plastic-scintillator array~\cite{NIMA.543.517}, with which neutron velocities are determined by measuring the ToF.
To achieve a sufficiently high ToF resolution, the detectors must be placed at a distance of up to \SI{15}{\m} from the target.
Detecting neutrons at forward angles also imposes a stringent constraint on the configuration of the magnets:
A dipole magnet to deflect the charged fragments away from the neutron path must have a large interpole gap so that neutrons can pass through.
Also, the subsequent magnets must be placed so that they do not obstruct the neutron path.
To enhance the acceptance for neutron detection, the target will be placed immediately in front of the dipole magnet (FS0$^\prime$ in Fig.~\ref{fig:layout}).

Most of these experiments will be carried out with neutron-rich rare isotopes of mass number of up to \num{132} and charge number up to \num{50}.
The accuracy of the invariant-mass reconstruction is dominated by the measurement accuracy of the neutron momentum vector.
The momentum-resolving-power requirement for the spectrometer is $1/\num{290} = 0.34\%$ (FWHM).
However, it is important that the momentum acceptance for charged fragments be large (\SI{\pm5}{\%}), because, especially in lighter neutron-rich systems, the decay by neutron emission induces significantly large momentum kicks which broaden the momentum distribution of the outgoing charged fragments.
A solid angle of \SI{10}{\msr} is necessary.

The ability to couple with auxiliary detectors is also important in this mode.
When using GRETA for invariant-mass-spectroscopy experiments, it is foreseen to remove its forward- as well as backward-detector modules to prevent obstructing of neutron path and relax space constraint, respectively.

Heavy-ion-collision experiments will be performed at the HRS using the S{\textpi}RIT-TPC (SAMURAI Pion Reconstruction and Ion-Tracker Time Projection Chamber)~\cite{NIMA.784.513}, which was originally developed for use in the SAMURAI dipole magnet at RIKEN RI Beam Factory (RIBF)~\cite{NIMB.317.294,IEEETransApplSupercon.23.4500308}.
Because the S{\textpi}RIT-TPC will be installed inside the dipole magnet, the beam transport is similar to that for neutron-invariant-mass experiments in that the reaction target is placed near the entrance of the dipole magnet.
The S{\textpi}RIT-TPC installation constrains the vertical gap size of the dipole magnet.
Based upon simulations using existing data from the RIBF experiments to assess the reconstruction efficiency and the quality of reconstruction of the reaction plane, it has been determined that a gap size of at least \SI{60}{\cm} be necessary;
a further increase would result in relatively small gains in efficiency and data-quality.

\subsection{HTBL}
\label{subsection:specificationsHTBL}

To enable and support the operation of the Spectrometer Section as described above, the HTBL must possess a number of key properties.
First and foremost, the HTBL must be able to optimally transport beams from ARIS by accommodating the maximum magnetic rigidity, \SI{8}{\Tm}, and the emittances of beams delivered to the HTBL.
The HTBL must also be able to achieve dispersion matching to the Spectrometer Section as discussed above.
The HTBL must have beam-tracking and diagnostics capabilities, which enable event-by-event measurement of beam momentum and beam-particle identification, which makes it possible to exploit cocktails of rare-isotope beams and study multiple reaction channels simultaneously.
Finally, the HTBL must be able to accommodate a radiofrequency fragment separator (RFFS).
The RFFS at NSCL~\cite{NIMA.606.314} has been proven useful for improving purities of very proton-rich beams, for which less proton-rich contaminants can otherwise attain levels that hinder experiments or render them unfeasible.
The HTBL must provide space and the ion-optical properties appropriate for the operation of an RFFS.

\section{Ion-optical design of the HTBL}
\label{section:ionOpticalDesignHTBL}

The HTBL is an analysis beamline starting at FB0, the final focal plane of ARIS, and ending at FS0, one of the two target locations of the Spectrometer Section (see Fig.~\ref{fig:layout}).
The length of the HTBL is \SI{41.70}{\meter}.
It consists of four \SI{22.5}{\degree} bending dipole magnets (denoted as DB1 through DB4 in Fig.~\ref{fig:layout}) and eight quadrupole triplets (denoted as TB1 through TB8 in Fig.~\ref{fig:layout}).
These magnets must be able to accommodate magnetic rigidities of up to \SI{8}{\Tm}.
The overall layout of the HTBL is mostly symmetric, which is beneficial for suppressing higher-order aberrations.
One minor modification from the perfect symmetry was necessary due to the space constraints around FB0 and FS0;
at FB0 the space is constrained by ARIS and the other beamlines, while at FS0 a larger space is needed for the installation of auxiliary detectors surrounding the reaction target.
Correction of higher-order aberrations is important in transporting large phase-space rare-isotope beams from ARIS.
Hardware correction of aberrations will be done by using sextupole and octupole corrector coils that are superimposed onto the quadrupole magnets.
Software correction in trajectory reconstruction can be used to further improve the precision in momentum and angle determination of beam particles delivered to the Spectrometer Section.

\subsection{Beam-transport modes}
\label{subsection:beamTransportModes}

\subsubsection{Realization of achromatic and dispersive transport modes}

\begin{figure}[!t]
    \centering
  \includegraphics{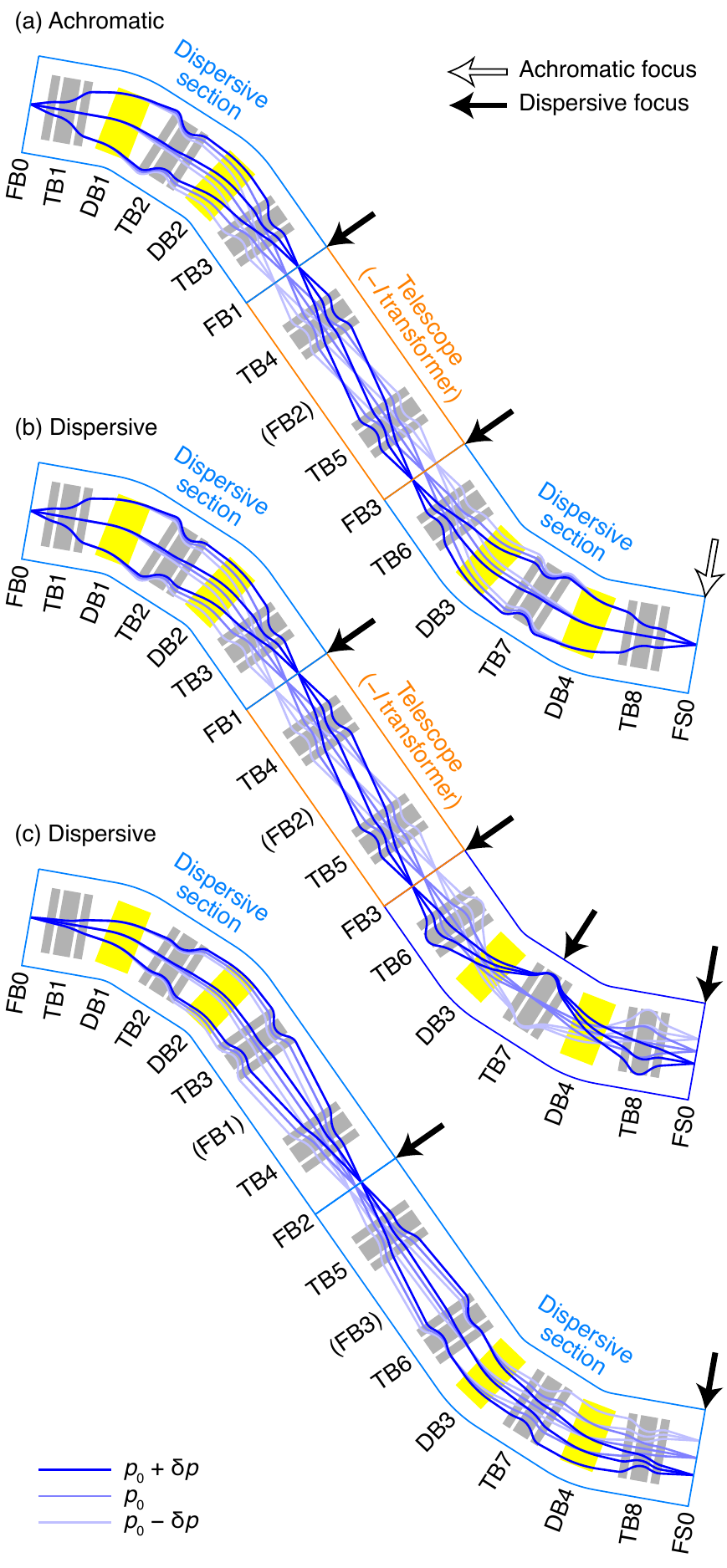}
  \caption{%
    These figures illustrate how the HTBL realizes achromatic and dispersive beam-transport modes.
    One concept for achromatic [(a)] and two for dispersive [(b) and (c)] beam-transport modes are shown by trajectories in the dispersive plane.
    See text for details.
  }
  \label{fig:beam_transport_mode_concepts}
\end{figure}

Figure~\ref{fig:beam_transport_mode_concepts} illustrates how the HTBL realizes achromatic and dispersive beam-transport modes. 

To make the HTBL achromatic, the horizontal dispersion created by the four dipole magnets must cancel out.
This is accomplished by implementing two intermediate foci between dipoles DB2 and DB3 in the dispersive plane as shown in Fig.~\ref{fig:beam_transport_mode_concepts}(a).
This makes two dispersive sections, one consisting of the magnets from TB1 to TB3 and the other of those from TB6 to TB8, connected by a non-dispersive telescope consisting of TB4 and TB5.
This telescope is often referred to as a ``$-I$ transformer'' because it inverts the signs of the position and angular magnifications without changing their magnitudes.
The first dispersive section makes a dispersive focus at FB1, and the telescope makes the next dispersive focus at FB3, where the sign of the dispersion is inverted.
The dispersion accumulated up to FB3 is canceled by the second dispersive section, and the target location FS0 becomes achromatic [$(x \vert \delta) = (x^\prime \vert \delta) = 0$].
At FS0, in addition to stigmatic, point-to-point focusing [$(x \vert x^\prime) = (y \vert y^\prime) = 0$], parallel-to-parallel imaging is achieved in both planes [$(x^\prime \vert x) = (y^\prime \vert y) = 0$].

To make the HTBL dispersive, on the other hand, the total dispersion created by the four dipole magnets must remain finite.
One way to accomplish this is to modify the section from TB6 to TB8 in the above-discussed achromatic tune such that dispersion accumulated up to FB3 will not be canceled out.
Shown in Fig.~\ref{fig:beam_transport_mode_concepts}(b) is a tune where an additional intermediate focus is implemented between DB3 and DB4 in the dispersive plane, by which the dispersion created by DB3 and that by DB4 cancel out partially or even fully, and the net dispersion of this section is much smaller than the net dispersion created by DB1 and DB2.
As a result, the dispersion accumulated up to FB3 is carried over by the remainder of the HTBL, and the final focus FS0 becomes a dispersive focus.
The partially-dispersion-matched beam transport discussed below is based upon this tune.

Another way to make the HTBL dispersive is to implement a single focus between DB2 and DB3 in the dispersive plane so that the dispersion created by the four dipole magnets simply accumulates as shown in Fig.~\ref{fig:beam_transport_mode_concepts}(c).
The upstream dispersive section (TB1 through TB4) and the downstream dispersive section (TB5 through TB8) are concatenated at the intermediate dispersive focus FB2, which makes the final focus FS0 also a dispersive focus.
The fully-dispersion-matched beam transport discussed below is based upon this tune.
The beam envelope must be expanded to gain dispersion by increasing the magnetic flux of the dipoles swept by the beam envelope (i.e.~for efficient illumination of the dipole fields).
The beam image in the dispersive plane at FS0 becomes stretched due to the resultant large $(x \vert \delta)$ value, and the angular and momentum acceptances are smaller compared to those of the achromatic tune.

\begin{figure*}[p]
  \captionof{table}{%
    Elements of the first-order transfer maps of the HTBL in the achromatic mode and in the partially- and fully-dispersion-matched modes.
    Those of the partially-dispersion-matched mode are identical to those of the achromatic mode up to FB3.
    Listed here are the ten non-trivial elements; the others are zero except $(\delta \vert \delta) = 1$.
  }
  \centering
  \begin{tabularx}{\textwidth}{lll*{9}{D{.}{.}{-1}}}
    \toprule
    & & & \multicolumn{3}{c}{Achromatic} 
        & \multicolumn{3}{c}{Partially-dispersion-matched} 
        & \multicolumn{3}{c}{Fully-dispersion-matched} \\
    \cmidrule(lr){4-6}
    \cmidrule(lr){7-9}
    \cmidrule(lr){10-12}
    & & & \multicolumn{1}{X}{\centering FB1} 
        & \multicolumn{1}{X}{\centering FB3}
        & \multicolumn{1}{X}{\centering FS0} 
        & \multicolumn{1}{X}{\centering FB1} 
        & \multicolumn{1}{X}{\centering FB3}
        & \multicolumn{1}{X}{\centering FS0} 
        & \multicolumn{1}{X}{\centering FB2}
        & \multicolumn{1}{X}{\centering FS0} \\
    \midrule
    $(x \vert x)$               &         & = $R_{11}$ & -1.44 &  1.44 & -1.19 & -1.44 &  1.44 &  1.07 & -0.58 &  0.35 \\
    $(x \vert x^\prime)$        & [m/rad] & = $R_{12}$ &  0.00 &  0.00 &  0.00 &  0.00 &  0.00 &  0.00 &  0.00 &  0.00 \\
    $(x^\prime \vert x)$        & [rad/m] & = $R_{21}$ &  0.00 &  0.00 &  0.00 &  0.00 &  0.00 & -0.27 & -0.38 &  0.62 \\
    $(x^\prime \vert x^\prime)$ &         & = $R_{22}$ & -0.69 &  0.69 & -0.84 & -0.69 &  0.69 &  0.94 & -1.71 &  2.83 \\
    $(x \vert \delta)$          & [m]     & = $R_{16}$ &  4.38 & -4.38 &  0.00 &  4.38 & -4.38 & -3.19 &  4.65 & -6.63 \\
    $(x^\prime \vert \delta)$   & [rad]   & = $R_{26}$ &  0.00 &  0.00 &  0.00 &  0.00 &  0.00 &  1.02 &  0.00 &  0.00 \\
    $(y \vert y)$               &         & = $R_{33}$ &  1.90 & -1.90 & -1.95 &  1.90 & -1.90 &  2.02 &  0.00 & -1.00 \\
    $(y \vert y^\prime)$        & [m/rad] & = $R_{34}$ &  0.00 &  0.00 &  0.00 &  0.00 &  0.00 &  0.00 &  2.04 &  0.00 \\
    $(y^\prime \vert y)$        & [rad/m] & = $R_{43}$ &  0.00 &  0.00 &  0.00 &  0.00 &  0.00 &  0.00 & -0.49 & -0.96 \\
    $(y^\prime \vert y^\prime)$ &         & = $R_{44}$ &  0.53 & -0.53 & -0.51 &  0.53 & -0.53 &  0.49 &  0.00 & -1.00 \\
    \bottomrule
  \end{tabularx}
  \label{table:HTBLmatrixElements}
\par
  \centering
  \includegraphics{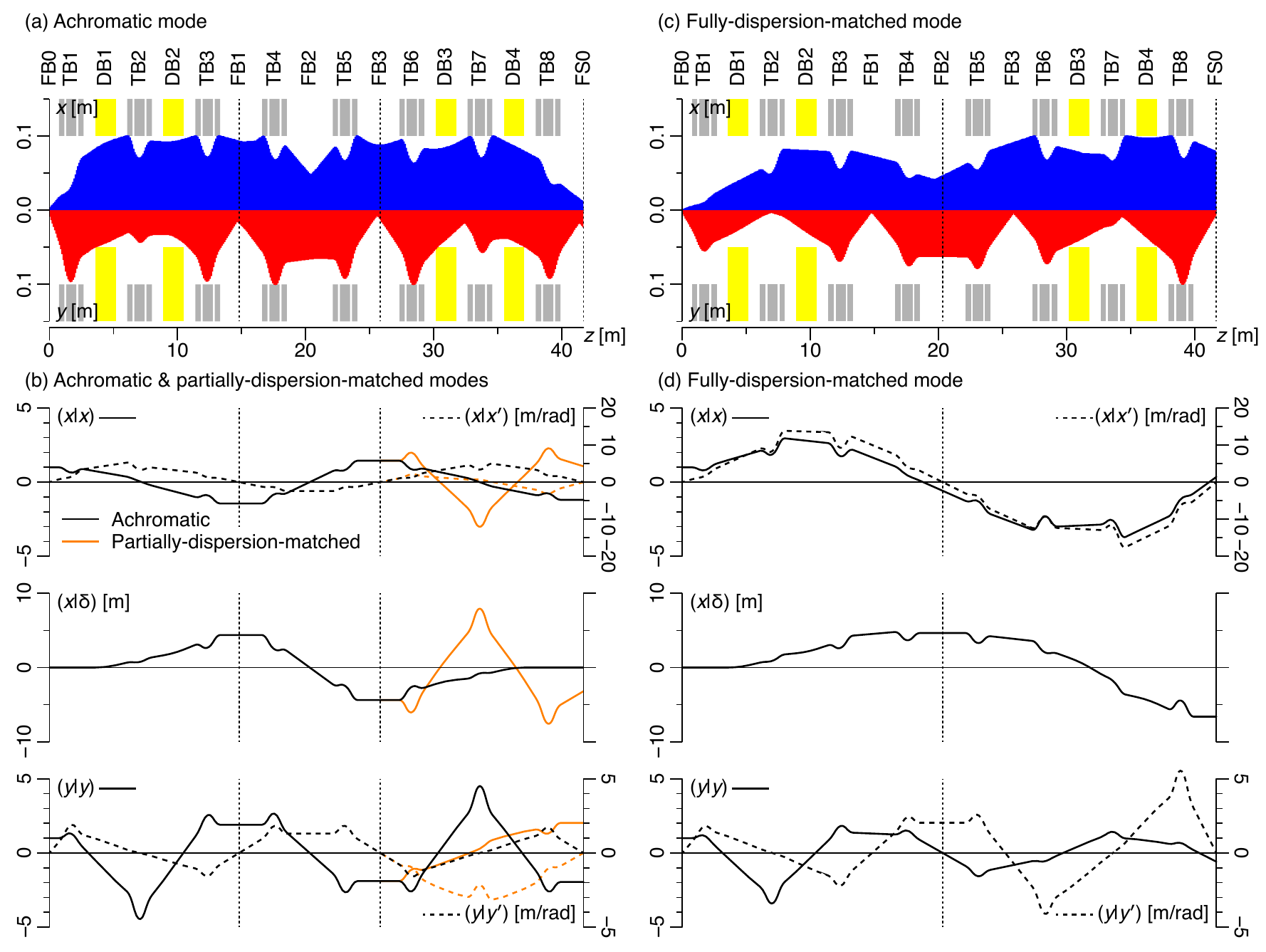}
    \captionof{figure}{%
    (a) Dispersive ($x$) and non-dispersive ($y$) fifth-order beam envelopes for the HTBL in the achromatic mode, depicted for the initial phase space of $\varDelta x^\prime_{\mathrm{max}} = \SI{20}{\mrad}$, $\varDelta y^\prime_{\mathrm{max}} = \SI{50}{\mrad}$, and $\varDelta p_{\mathrm{max}}/p = \SI{2}{\%}$.
    The dipole magnets are indicated by the yellow boxes, showing their good-field regions (\SI{\pm10}{\cm}) and vertical gaps (\SI{\pm5}{\cm}).
    The quadrupoles are indicated by the gray boxes, showing warm-bore aperture sizes (\SI{10}{\cm} in radius).
    (b) Cosine-like [$(x \vert x)$] and sine-like [$(x \vert x^\prime)$] functions in the dispersive plane (top), the dispersion [$(x \vert \delta)$] (middle), and the cosine-like [$(y \vert y)$] and sine-like [$(y \vert y^\prime)$] functions in the non-dispersive plane (bottom) for the HTBL in the achromatic mode (solid curves) and in the partially-dispersion-matched mode (dotted curves). These are identical up to FB3.
    (c) Same as (a) but for the fully-dispersion-matched mode.
    The initial phase space is $\varDelta x^\prime_{\mathrm{max}} = \SI{6}{\mrad}$, $\varDelta y^\prime_{\mathrm{max}} = \SI{30}{\mrad}$, and $\varDelta p_{\mathrm{max}}/p = \SI{1}{\%}$.
    (d) Same as (b) but for the fully-dispersion-matched mode.
  }
  \label{fig:HTBL_envelopes_MEs}
\end{figure*}

\subsubsection{Achromatic beam-transport mode}

The first-order transfer-map elements are listed in Table~\ref{table:HTBLmatrixElements}, and the first-order ion-optical plots are shown in Fig.~\ref{fig:HTBL_envelopes_MEs}.
Note that the position magnifications from FB0 to FS0 in the dispersive and non-dispersive planes are larger than one, which in turn makes the angular magnifications, the inverse of the position magnifications, smaller than one.
This comes from the fact that the space around FS0 is stretched compared to that around FB0.

At the location of the dispersive foci FB1 and FB3, the momenta of beam particles can be determined by measuring their dispersive positions.
Likewise, by measuring angles at FB1 or FB3, the incoming beam angle at FS0 can be inferred.
Such momentum or angle measurements are necessary when a better resolution is needed in the reaction analysis in the spectrometer than the momentum or angular spread of the incoming beam at FS0.

\subsubsection{Dispersion-matched beam-transport mode}

The first-order transfer-map elements are listed in Table~\ref{table:HTBLmatrixElements}, and the first-order ion-optical plots are shown in Fig.~\ref{fig:HTBL_envelopes_MEs}.

For full dispersion matching, the lateral and angular dispersions of the HTBL are matched to those of the Spectrometer Section from FS0 to FS2, making the HRS from FB0 to FS2 achromatic. In case of partial dispersion matching, the HTBL dispersions are matched to the Spectrometer Section from FS0 to FS1.
To achieve dispersion matching, the lateral and angular dispersions of the HTBL need to satisfy the conditions imposed by the ion-optical properties of the Spectrometer Section.
The condition for the lateral dispersion is
\begin{subequations}
  \begin{equation}
    (x \vert \delta)_{\mathrm{B}} = - \frac{(x \vert \delta)_{\mathrm{S}}}{(x \vert x)_{\mathrm{S}}},
  \end{equation}
and that for the angular dispersion is 
  \begin{equation} \label{eq:angularDispersion}
    (x^\prime \vert \delta)_{\mathrm{B}} = (x^\prime \vert x)_{\mathrm{S}} (x \vert \delta)_{\mathrm{S}} - (x \vert x)_{\mathrm{S}} (x^\prime \vert \delta)_{\mathrm{S}},
  \end{equation}
\end{subequations}
where the subscripts B and S denote the beamline (HTBL) and the spectrometer (Spectrometer Section), respectively (more details are given in \ref{Appendix:DM}).
In the current design of the Spectrometer Section, $(x^\prime \vert x)_{\mathrm{S}}$ and $(x^\prime \vert \delta)_{\mathrm{S}}$ are set to zero, which zeroes RHS of Eq.~\eqref{eq:angularDispersion}, requiring that the angular dispersion of HTBL vanish for the dispersion matching.

\subsection{Accommodation of a Radiofrequency Fragment Separator (RFFS)}

A conceptual design of an RFFS to be used in the HTBL that provides the necessary level of contaminant reduction has been developed~\cite{EPJTI.7.4} based upon a double quarter-wave resonator cavity~\cite{NIMA.234.235}.
The electrode plates are \SI{1.86}{\meter} long, and the gap between them is \SI{18}{\cm}, as is necessary to transmit large emittance rare-isotope beams.
It operates at a frequency of \SI{20.125}{\MHz} and has a maximum voltage of \SI{\pm182}{\kV} or a maximum gradient of \SI{20.2}{\kV/\cm}.
The total flange-to-flange length of this RFFS is \SI{2.20}{\meter} and the total height is \SI{3.72}{\meter}.
Simulations demonstrated that the angular deflection of \SI{\pm8.6}{\mrad} can be achieved for ${}^{100}\mathrm{Sn}^{50+}$ at \SI{100}{\MeV}/$u$.

In all the beam-transport modes presented above, the beam becomes parallel in the non-dispersive plane [$(y^\prime \vert y^\prime) = 0$] at FB2, which makes it a suitable location to install a vertically deflecting RFFS.
For its operation, vertical steering magnets are additionally needed immediately before and after the RFFS to keep the particles of interest centered around the beam axis for further transmission in the HTBL.
Variable slits to reject deflected contaminants and detectors to diagnose the filtering by the RFFS will be installed at FB3.

\begin{figure}[t]
  \centering
  \includegraphics{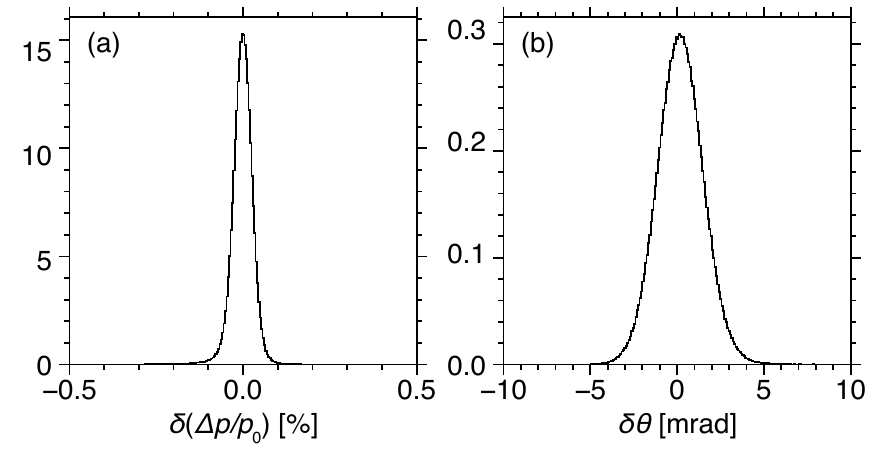}
  \caption{%
    (a) Momentum resolution in the achromatic mode calculated as the difference between the initial and reconstructed momenta in trajectory-reconstruction simulations using COSY-calculated maps including up to fifth order.
    (b) Same as (a) but for the angular resolution.
  }
  \label{fig:HTBL_resolutions}
\end{figure}

\subsection{Momentum and angular resolutions}
\label{subsection:ResolutionsHTBL}

The momentum and angular resolutions in the achromatic mode were evaluated by means of Monte-Carlo simulations using fifth-order COSY maps.
A ${}^{40}\mathrm{Mg}$ beam produced in ARIS from a ${}^{48}\mathrm{Ca}$ primary beam with a carbon production target was used for the simulations as a representative beam having a large phase space at FB0, where the beam-spot size was estimated to be \SI{1.4}{\mm} ($x$) $\times$ \SI{4.0}{\mm} ($y$) in FWHM.

Beam particles were transported from FB0 to the dispersive focus FB3, and the coordinates were obtained using the COSY map $\mathcal{M}$ which relates the coordinates at FB0 and those at FB3 as
\begin{subequations}
\begin{equation} \label{eq:forwardMapping}
  \begin{pmatrix}
    x _{\mathrm{FB3}} \\
    a _{\mathrm{FB3}} \\
    y _{\mathrm{FB3}} \\
    b _{\mathrm{FB3}} \\
    d _{\mathrm{FB3}}
  \end{pmatrix}
  = \mathcal{M}
  \begin{pmatrix}
    x _{\mathrm{FB0}} \\
    a _{\mathrm{FB0}} \\
    y _{\mathrm{FB0}} \\
    b _{\mathrm{FB0}} \\
    d _{\mathrm{FB0}}
  \end{pmatrix}
\end{equation}
in COSY notation~\cite{Berz.AIEP108book}.
Here $d _{\mathrm{FB3}} = d _{\mathrm{FB0}}$ because no momentum change occurs between the two locations.
Following Ref.~\cite{PhysRevC.47.537}, under the assumption that the initial position $x _{\mathrm{FB0}}$ is small and therefore approximated by zero, Eq.~\eqref{eq:forwardMapping} can be reduced with a nonlinear submap $\mathcal{S}$ to
\begin{equation}
  \begin{pmatrix}
    x _{\mathrm{FB3}} \\
    a _{\mathrm{FB3}} \\
    y _{\mathrm{FB3}} \\
    b _{\mathrm{FB3}}
  \end{pmatrix}
  = \mathcal{S}
  \begin{pmatrix}
    a _{\mathrm{FB0}} \\
    y _{\mathrm{FB0}} \\
    b _{\mathrm{FB0}} \\
    d _{\mathrm{FB0}}
  \end{pmatrix},
\end{equation}
\end{subequations}
where $\mathcal{S}$ is invertible to arbitrary order.
The momentum was reconstructed from the measured positions and angles at FB3 using the inverse map $\mathcal{S}^{-1}$ from FB3 to FB0,
\begin{equation} \label{eq:inverseMapping}
  \begin{pmatrix}
    a _{\mathrm{FB0}} \\
    y _{\mathrm{FB0}} \\
    b _{\mathrm{FB0}} \\
    d _{\mathrm{FB0}}
  \end{pmatrix} _{\mathrm{recon}}
  = \mathcal{S} ^{-1}
  \begin{pmatrix}
    x _{\mathrm{FB3}} \\
    a _{\mathrm{FB3}} \\
    y _{\mathrm{FB3}} \\
    b _{\mathrm{FB3}}
  \end{pmatrix} _{\mathrm{meas}},
\end{equation}
where the subscripts $_{\mathrm{recon}}$ and $_{\mathrm{meas}}$ stand for the reconstructed and measured quantities, respectively.

Position and angle measurements were simulated at FB3 with a pair of tracking detectors. Randomized errors were added to the positions to model the finite position resolution of the tracking detectors which was assumed to be \SI{1}{\mm} (FWHM).
The accuracy of the reconstruction is limited by the validity of the above-mentioned assumption that $x _{\mathrm{FB0}} = 0$ as well as the detector resolution.
Here, the effects of higher-order aberrations were included by using maps calculated up to fifth order.
The momentum resolution calculated as the difference between the initial (input) and reconstructed momenta was estimated to be $1/\num{1700} = 0.06\%$ (FWHM) as shown in Fig.~\ref{fig:HTBL_resolutions}(a).
The angle at FS0 was inferred using the positions and angles at FB3 and the transfer map from FB3 to FS0, and the angular resolution was obtained from the difference between this inferred angle and the angle directly calculated from the transfer map from FB0 to FS0.
The angular resolution was \SI{3.1}{\mrad} (FWHM) as shown in Fig.~\ref{fig:HTBL_resolutions}(b).

\begin{figure}[!t]
  \centering
  \includegraphics{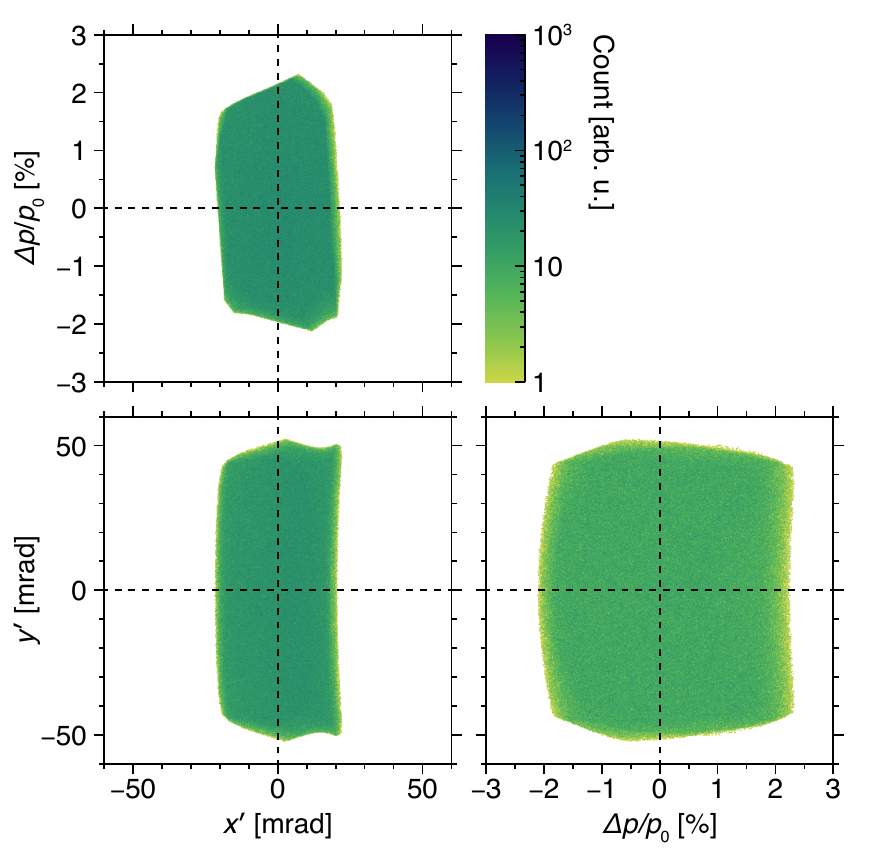}
  \caption{%
    Momentum and angular acceptances for the achromatic mode of the HTBL as simulated in LISE$^{++}_{\mathit{cute}}$ by using the fifth-order COSY maps.
    The aperture sizes of the quadrupole magnets, the good-field regions in the dispersive plane, and the vertical interpole gap sizes in the non-dispersive plane of the dipole magnets were taken into account.
  }
  \label{fig:HTBL_HA_acceptances}
\end{figure}

\begin{figure}[!t]
  \centering
  \includegraphics{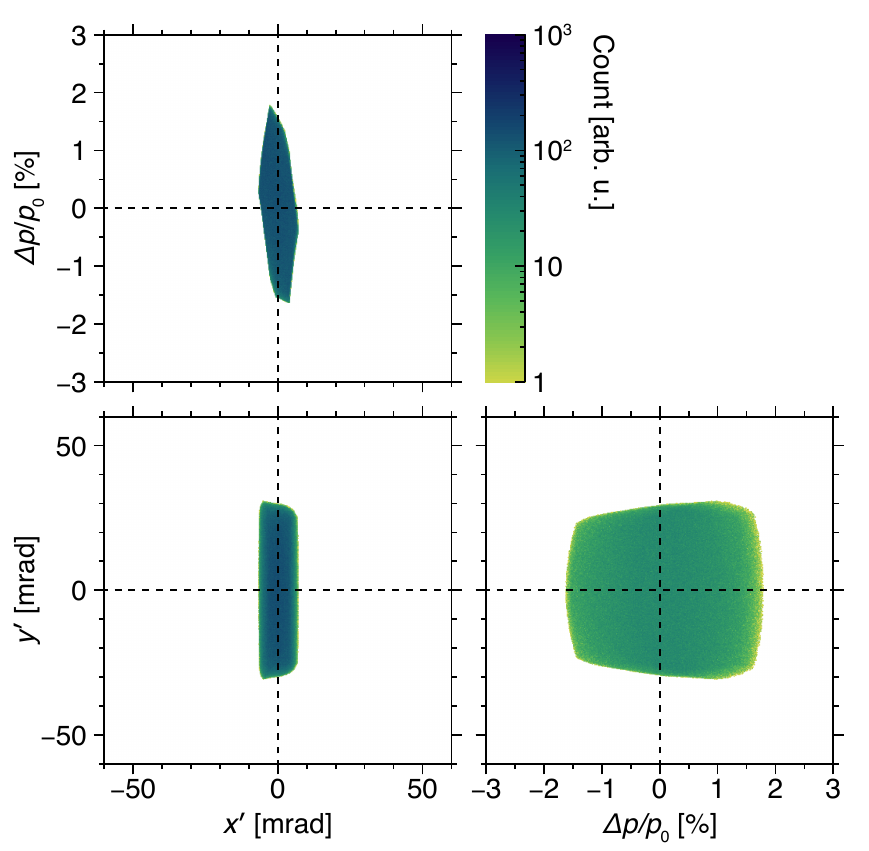}
  \caption{%
    Momentum and angular acceptances for the fully-dispersion-matched mode as simulated in LISE$^{++}_{\mathit{cute}}$ by using the fifth-order COSY maps.
    The aperture sizes of the quadrupole magnets, the good-field regions in the dispersive plane, and the vertical interpole gap sizes in the non-dispersive plane of the dipole magnets were taken into account.
  }
  \label{fig:HTBL_DM_acceptances}
\end{figure}
  
\begin{figure}[!t]
  \centering
  \includegraphics{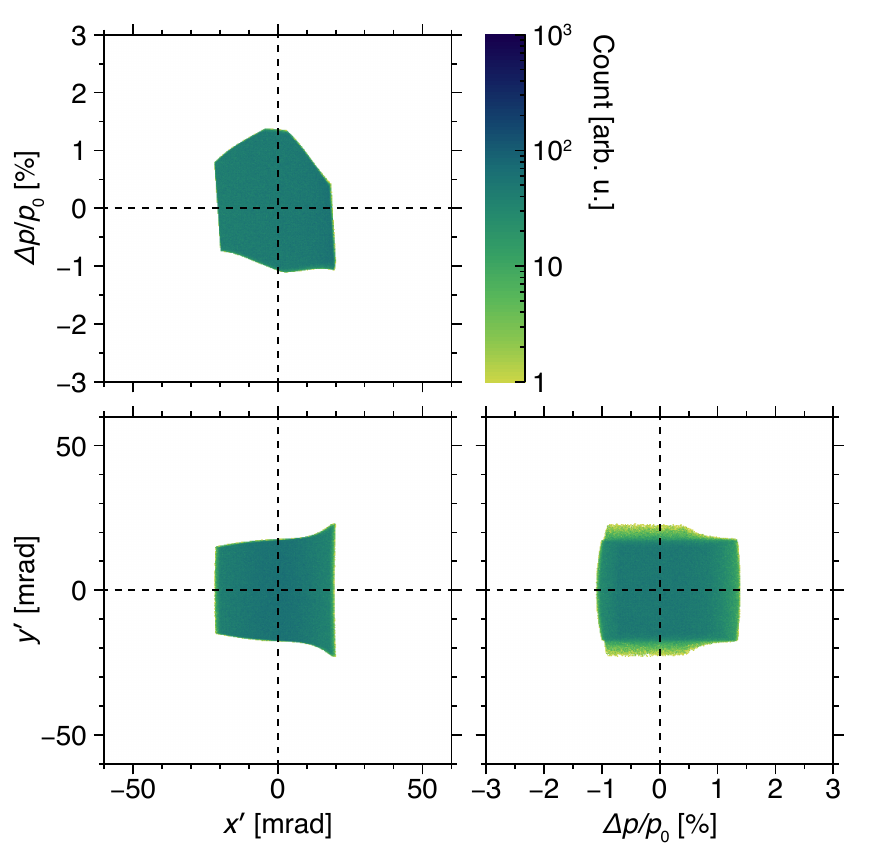}
  \caption{%
    Momentum and angular acceptances for the partially-dispersion-matched mode as simulated in LISE$^{++}_{\mathit{cute}}$ by using the fifth-order COSY maps.
    The aperture sizes of the quadrupole magnets, the good-field regions in the dispersive plane, and the vertical interpole gap sizes in the non-dispersive plane of the dipole magnets were taken into account.
  }
  \label{fig:HTBL_pDM_acceptances}
\end{figure}

\subsection{Momentum and angular acceptances}
\label{subsection:AcceptancesHTBL}

The momentum and angular acceptances were also evaluated for all the three beam-transport modes by means of Monte-Carlo simulations using LISE$^{++}_{\mathit{cute}}$ with fifth-order COSY maps.
The aperture sizes of the quadrupole magnets, the good-field regions in the dispersive plane, and the vertical interpole gap sizes in the non-dispersive plane of the dipole magnets were taken into account.
Transmitted from a point source at FB0, beam particles with uniform momentum and angular distributions were injected into the HTBL.
For the particles that reached at FS0, three two-dimensional correlations between the phase-space variables at FB0 were plotted in Figs.~\ref{fig:HTBL_HA_acceptances}, \ref{fig:HTBL_DM_acceptances}, and \ref{fig:HTBL_pDM_acceptances}, which represent the momentum and angular acceptances of each beam-transport mode.
In the achromatic mode (Fig.~\ref{fig:HTBL_HA_acceptances}), the angular acceptances are 
$\lvert \varDelta x^\prime \rvert \lesssim \SI{20}{\mrad}$
and 
$\lvert \varDelta y^\prime \rvert \lesssim \SI{50}{\mrad}$, 
and the momentum acceptance is
$\lvert \varDelta p/p \rvert \lesssim \SI{2}{\%}$,
in the fully-dispersion-matched mode (Fig.~\ref{fig:HTBL_DM_acceptances}), they are 
$\lvert \varDelta x^\prime \rvert \lesssim \SI{6}{\mrad}$
and
$\lvert \varDelta y^\prime \rvert \lesssim \SI{30}{\mrad}$,
and
$\lvert \varDelta p/p \rvert \lesssim \SI{1.5}{\%}$,
and in the partially-dispersion-matched mode (Fig.~\ref{fig:HTBL_pDM_acceptances}), they are 
$\lvert \varDelta x^\prime \rvert \lesssim \SI{20}{\mrad}$
and
$\lvert \varDelta y^\prime \rvert \lesssim \SI{20}{\mrad}$,
and
$\lvert \varDelta p/p \rvert \lesssim \SI{1}{\%}$.
Note that in the fully-dispersion-matched mode the angular and momentum acceptances is smaller compared to those of the achromatic mode, because the beam envelope is expanded for the optimal illumination of the dipole fields.
Also note that in the partially-dispersion-matched mode the non-dispersive-angular acceptance is smaller than the other two modes due to the modification of the beam tune after FB3, where the node in the non-dispersive plane between DB3 and DB4 was eliminated to implement one in the dispersive plane instead, resulting in a broader beam envelope in the non-dispersive plane.

\section{Ion-optical design of the Spectrometer Section}
\label{section:ionOpticalDesignSpectrometerSection}

\begin{figure}[!t]
  \centering
  \includegraphics{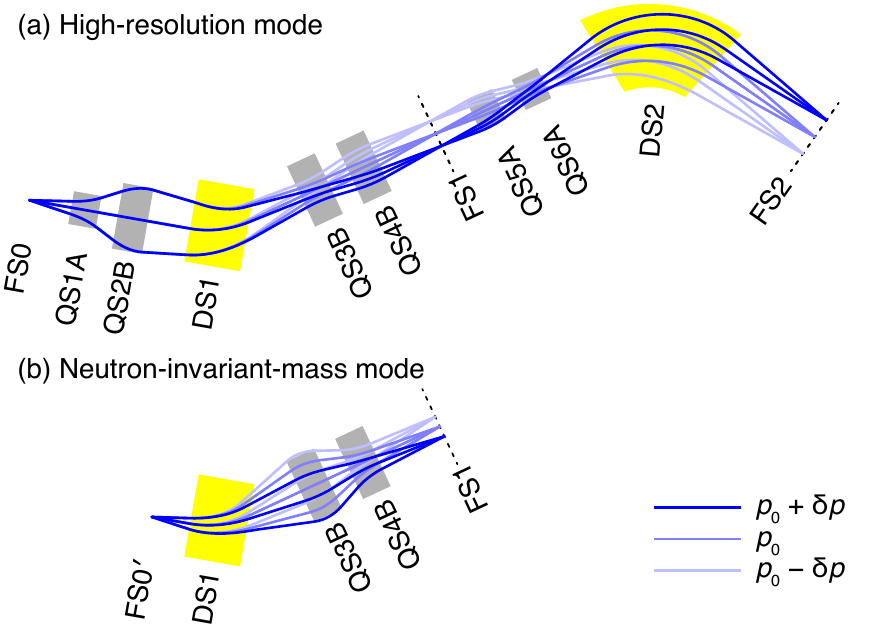}
  \caption{%
  Ion-optical design concepts are shown by trajectories in the dispersive plane for (a) the high-resolution mode and (b) the neutron-invariant-mass mode of the Spectrometer Section.
    See text for details.
  }
  \label{fig:SpectrometerSection_ion_optics_concepts}
\end{figure}

The Spectrometer Section starts at FS0, the end of the HTBL, and its total length is \SI{30.80}{\meter} (see Fig.~\ref{fig:layout}).
The Spectrometer Section has a \SI{35}{\degree} left-bending dipole magnet (denoted as DS1 in Fig.~\ref{fig:layout}), a \SI{60}{\degree} right-bending dipole magnet (denoted as DS2 in Fig.~\ref{fig:layout}), and six quadrupoles of two types, all with superimposed sextupole and octupole corrector coils.
These magnets are designed such that they accommodate magnetic rigidities up to \SI{8}{\Tm}.
These magnets have large aperture sizes, which enables to achieve large angular and momentum acceptances while affording large space around the reaction target, which is needed for installation of a variety of auxiliary detectors.
Optimizations of the ion-optical layout can be best achieved by making the first two quadrupoles, QS1A and QS2B, movable along the beam direction.
Such a moving capability is being explored, and all the figures and calculations presented in this paper are based on the lattice layout with the movable QS1A and QS2B.

Figure~\ref{fig:SpectrometerSection_ion_optics_concepts} illustrates the concepts of the ion-optical design of the Spectrometer Section, which is described in the following in more detail for each operational mode.

\begin{figure*}[p]
  \captionof{table}{%
    Elements of the first-order transfer maps in the Spectrometer Section.
    Listed here are the ten non-trivial elements; the others are zero except $(\delta \vert \delta) = 1$.
  }
  \centering
  \begin{tabularx}{\textwidth}{lll*{5}{D{.}{.}{-1}}}
    \toprule
    & & & \multicolumn{2}{c}{High-resolution mode} 
        & \multicolumn{1}{c}{Neutron-invariant-mass mode} \\
    \cmidrule(lr){4-5}
    \cmidrule(lr){6-6}
    & & & \multicolumn{1}{X}{\centering FS1} 
        & \multicolumn{1}{X}{\centering FS2} 
        & \multicolumn{1}{c}{\centering FS1} \\
    \midrule
    $(x \vert x)$               &         & = $R_{11}$ & -2.03 &  1.49  & -1.46 \\
    $(x \vert x^\prime)$        & [m/rad] & = $R_{12}$ &  0.00 &  0.00  &  0.00 \\
    $(x^\prime \vert x)$        & [rad/m] & = $R_{21}$ & -0.16 &  0.00  & -0.47 \\
    $(x^\prime \vert x^\prime)$ &         & = $R_{22}$ & -0.49 &  0.67  & -0.69 \\
    $(x \vert \delta)$          & [m]     & = $R_{16}$ & -6.47 &  9.86  & -2.16 \\
    $(x^\prime \vert \delta)$   & [rad]   & = $R_{26}$ &  0.00 &  0.00  & -0.30 \\
    $(y \vert y)$               &         & = $R_{33}$ &  4.73 &  0.00  &  0.00 \\
    $(y \vert y^\prime)$        & [m/rad] & = $R_{34}$ & -0.47 &  0.81  &  2.49 \\
    $(y^\prime \vert y)$        & [rad/m] & = $R_{43}$ &  2.13 & -1.23  & -0.40 \\
    $(y^\prime \vert y^\prime)$ &         & = $R_{44}$ &  0.00 &  0.15  & -3.14 \\
      \bottomrule
    \end{tabularx}
  \label{table:HRmatrixElements}
\par
  \centering
  \includegraphics{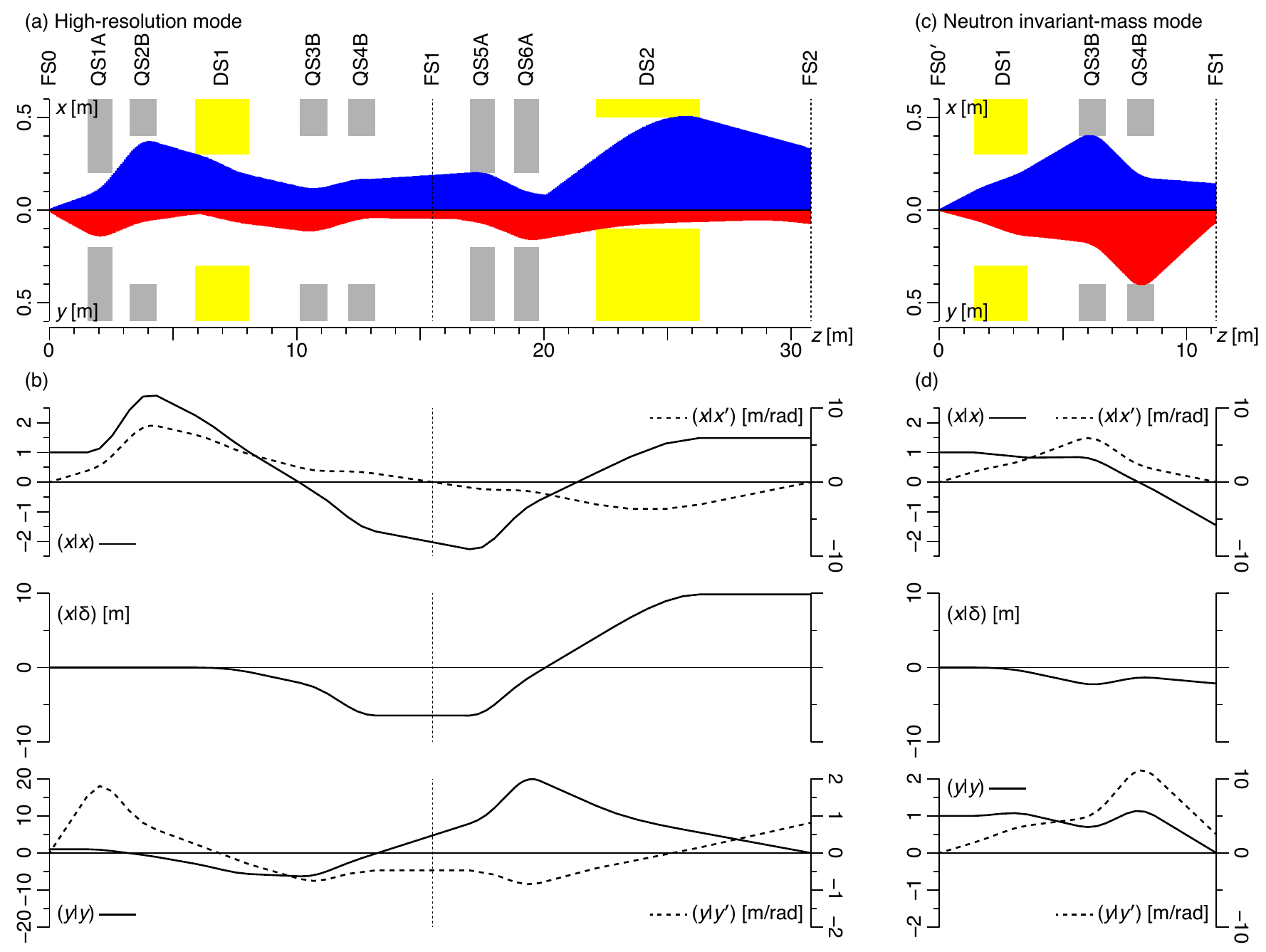}
  \captionof{figure}{%
    (a) Dispersive ($x$) and non-dispersive ($y$) fifth-order beam envelopes for the Spectrometer Section in the high-resolution mode, depicted for the initial phase space of $\varDelta x^\prime_{\mathrm{max}} = \SI{50}{\mrad}$, $\varDelta y^\prime_{\mathrm{max}} = \SI{75}{\mrad}$, and $\varDelta p_{\mathrm{max}}/p = \SI{2.5}{\%}$. The dipole magnets are indicated by the yellow boxes and the quadrupoles are indicated by the gray boxes.
    (b) Cosine-like [$(x \vert x)$] and sine-like [$(x \vert x^\prime)$] functions in the dispersive plane (top), the dispersion [$(x \vert \delta)$] (middle), and the cosine-like [$(y \vert y)$] and sine-like [$(y \vert y^\prime)$] functions in the non-dispersive plane (bottom) for the Spectrometer Section in the high-resolution mode.
    (c) Same as (a) but for the neutron-invariant-mass mode, depicted for the initial phase space of $\varDelta x^\prime_{\mathrm{max}} = \SI{70}{\mrad}$, $\varDelta y^\prime_{\mathrm{max}} = \SI{36}{\mrad}$, and $\varDelta p_{\mathrm{max}}/p = \SI{5}{\%}$.
    (d) Same as (b) but for the neutron-invariant-mass mode.
  }
  \label{fig:SS_envelopes_MEs}
\end{figure*}

\subsection{High-resolution mode}

\subsubsection{Ion-optical design}
\label{subsection:ionOpticalDesignHighResolutionMode}

The high-resolution mode utilizes all the elements of the Spectrometer Section from FS0 to FS2.
QS1A expands the beam in the dispersive plane for optimal dipole illumination of DS1, while it collapses the beam in the non-dispersive plane.
QS2B, with a larger bore size, has the opposite polarity and optimizes the transmission through DS1.
DS1 is a rectangular dipole.
The entrance pole face is perpendicular to the beam axis.
The exit pole face in turn is rotated by \SI{35}{\degree}, which provides horizontal defocusing and vertical focusing.
QS3B and QS4B capture the beam exiting from DS1 and guide them through the rest of the Spectrometer Section.
These quadrupoles are placed sufficiently far downstream of DS1 such that they do not obstruct the path of neutrons for the neutron-invariant-mass mode as described below.
An intermediate focus at FS1 is created so that the dispersions generated by the left-bending DS1 and the subsequent right-bending DS2 can be accumulated.
QS5A and QS6A transport the beam to DS2.
In the dispersive plane, the beam is expanded to illuminate DS2 optimally, while in the vertical direction a node is created inside DS2 to optimize the transmission.
DS2 has pole-face rotations of \SI{20}{\degree} at the entrance and the exit such that they provide horizontal focusing and vertical defocusing, shaping the beam envelope through the interpole gap towards the final focus FS2.

The first-order transfer-map elements are listed in Table~\ref{table:HRmatrixElements}, and the first-order ion-optical plots are shown in Fig.~\ref{fig:SS_envelopes_MEs}.
The six quadrupoles provide six degrees of freedom, which are used to achieve $(x \vert x^\prime) = (y^\prime \vert y^\prime) = 0$ at FS1, $(x \vert x^\prime) = (y \vert y) = (x^\prime \vert x) = (x^\prime \vert \delta) = 0$ at FS2.
The last two conditions are required to achieve the previously-mentioned zero angular dispersion of the HTBL in full dispersion matching.

The sextupole and octupole corrector coils superimposed onto all the quadrupoles are used to optimize the acceptance of the spectrometer.
This is important especially given the large magnet bore sizes.
It is also important that aberrations be corrected with these higher-order coils for the purpose of blocking unreacted beams, because aberrations will increase the beam-spot size that needs to be masked.
Momentum and scattering-angle reconstruction will be done using positions and angles measured at the focal plane with inverse maps including higher orders~\cite{PhysRevC.47.537}, as has been successfully done for analyses of data from experiments performed at the S800~\cite{NIMB.204.629}.

\subsubsection{Momentum and angular resolutions}

\begin{figure}[!t]
  \centering
  \includegraphics{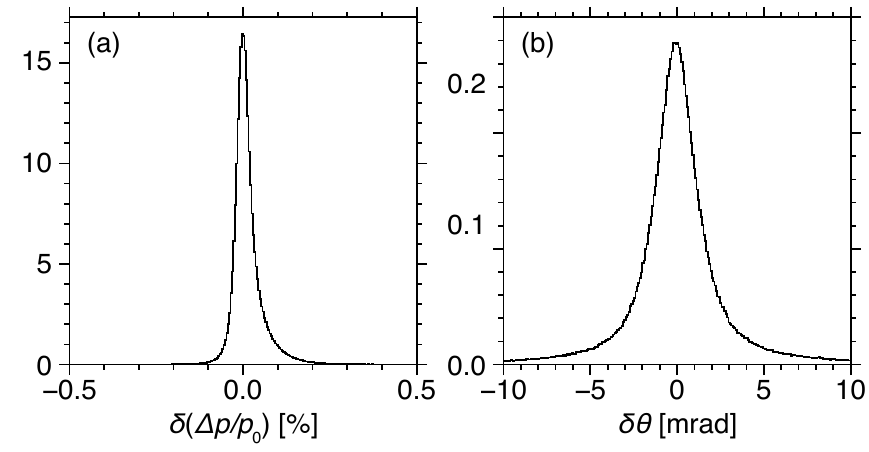}
  \caption{%
    (a) Momentum resolution in the high-resolution mode calculated as the difference between the initial and reconstructed momenta in trajectory-reconstruction simulations using COSY maps including up to fifth order.
    (b) Same as (a) but the angular resolution.
  }
  \label{fig:SS_HR_resolutions}
\end{figure}

The momentum and angular resolutions were evaluated by means of Monte-Carlo simulations using fifth-order COSY maps. The simulations are similar to those for the HTBL described in Section~\ref{subsection:ResolutionsHTBL}.
Particles emitted from FS0, where the initial beam-spot size was \SI{2.8}{\mm} ($x$) $\times$ \SI{8.2}{\mm} ($y$) in FWHM as estimated for a ${}^{40}\mathrm{Mg}$ beam delivered through the HTBL, were transported using the transfer maps from FS0 to the final focus FS2.
There, position and angle measurements with a pair of tracking detectors were simulated, where randomized errors were added to model the finite position resolution of the tracking detectors which was assumed to be \SI{1}{\mm} (FWHM).
Subsequently, the particles were traced backward from FS2 to FS0 via the inverse map to reconstruct the momentum and the scattering angle.
Figure~\ref{fig:SS_HR_resolutions} shows the momentum and angular resolutions, calculated as the difference between the initial and reconstructed momenta and angles.
The momentum and angular resolutions were estimated to be $1/\num{2000} = 0.05\%$ (FWHM) and \SI{2.7}{\mrad} (FWHM), respectively.

For reaction analysis, it is of greater interest to know the resolutions with which the momentum change (excitation energy) and angle changes (scattering angle) can be determined, and they were also evaluated.
A ${}^{204}\mathrm{Pt}$ beam produced from a ${}^{208}\mathrm{Pb}$ beam with a carbon production target in ARIS was used, and its beam-spot size at FB0 was \SI{1.2}{\mm} ($x$) $\times$ \SI{3.4}{\mm} ($y$) in FWHM, and the momentum spread of this beam was \SI{\pm0.3}{\%} in FWHM.
The beam was transported either in the fully-dispersion-matched mode or the achromatic mode in the HTBL to the reaction target at FS0.
At the target, changes in momentum [$\delta (\varDelta p/p)_{\mathrm{sc}}$] and angle ($\delta x^\prime_{\mathrm{sc}}$ and $\delta y^\prime_{\mathrm{sc}}$) were added to model the momentum transfer in the reaction.
The reaction product was then transported to FS2, where position and angle measurements with tracking detectors with a position resolution of \SI{1}{\mm} (FWHM) were simulated.
The momentum and the scattering angle at the target were then reconstructed in the same way as above.

\begin{figure}[!t]
\centering
  \includegraphics{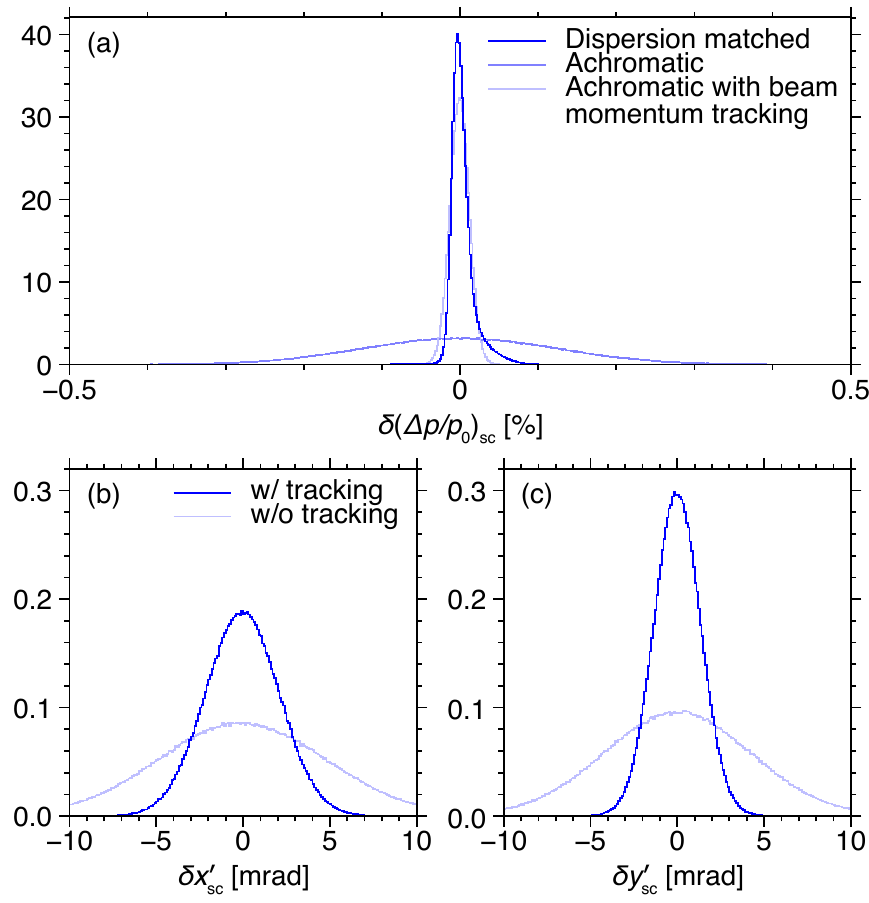}
  \caption{%
    (a) Resolutions of the determination of the momentum change by the reaction in the high-resolution mode. 
    The beam was transported either in the fully-dispersion-matched mode or in the achromatic mode in the HTBL.
    For the latter, those with and without tracking of the beam momentum at the dispersive focus FB3 are compared.
    (b) Resolutions of the determination of the dispersive component of the angle change by the reaction in the high-resolution mode.
    The beam is transported in the fully-dispersion-matched mode in the HTBL.
    Those with or without tracking of the incoming beam angle at the target are compared.
    (c) Same as (b) but for the non-dispersive component.
  }
  \label{fig:SS_HR_resolutions_matched}
\end{figure}

Figure~\ref{fig:SS_HR_resolutions_matched}(a) compares the resolutions of the determination of the momentum change.
When the beam was transported in the achromatic mode of the HTBL, the resolution was $1/\num{340} = 0.3\%$ (FWHM) which was dominated by the momentum spread of the beam (\SI{\pm0.3}{\%} in FWHM).
Tracking of the beam momentum at FB3 with a pair of tracking detectors with a position resolution of \SI{1}{\mm} (FWHM) improved the resolution to be $1/\num{3400} = 0.03\%$.
Note that this value is better than the above-mentioned momentum resolution of 1/\num{2000} (FWHM), which is owing to the smaller beam-spot size for the ${}^{204}\mathrm{Pt}$ beam than the ${}^{40}\mathrm{Mg}$ beam at FS0 in the achromatic mode, which improves the accuracy of the momentum reconstruction in which the particle is assumed to be emitted from $x_{\mathrm{FS0}} = \SI{0}{\mm}$.
When the beam was transported in the fully-dispersion-matched mode, a resolution of $1/\num{5000} = 0.02\%$ (FWHM) was achieved.
Again, this value is also better than 1/\num{2000} (FWHM) owing to the smaller monochromatic beam-spot size at FS0 which stems from the small position magnification [$(x \vert x) = 0.35$] in the fully-dispersion-matched mode.

Figure~\ref{fig:SS_HR_resolutions_matched}(b) and (c) compare the resolutions of determination of the dispersive and non-dispersive components of the angle changes, respectively.
With tracking of the incoming beam angles, the achieved resolutions were \SI{5.0}{\mrad} and \SI{3.2}{\mrad} (FWHM), to which the intrinsic resolutions of the angle determination, \SI{4.1}{\mrad} and \SI{2.3}{\mrad} (FWHM), and the resolution of the incoming beam angle measurement, \SI{2.9}{\mrad} (FWHM), both contributed. 
Without tracking of the incoming beam angles, the achievable resolutions were dominated by the spread of the incoming beam angles of about \SI{10}{\mrad} in both dispersive and non-dispersive planes.

These simulations demonstrated that dispersion matching makes it possible to achieve high precision in excitation-energy determination without beam momentum tracking, which obviates tracking of the beam momentum.
On the other hand, tracking of the incoming beam angle is necessary to deduce the scattering angle when it is necessary to measure the scattering angle with better precision than the angular spread of the beam, because the scattering angle is to be determined by taking the difference between the incoming angle measured before the target and the outgoing angle reconstructed from the positions and angles measured at the spectrometer focal plane.

\begin{figure}[t]
  \centering
  \includegraphics{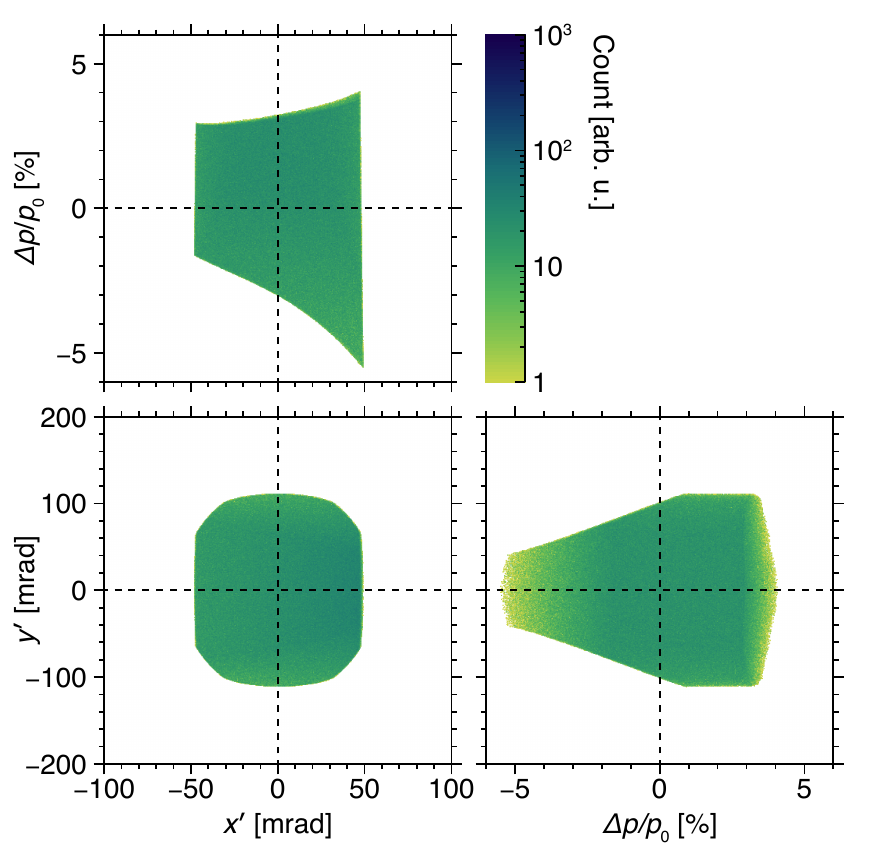}
  \caption{%
    Momentum and angular acceptances for the high-resolution mode of the Spectrometer Section as simulated in LISE$^{++}_{\mathit{cute}}$ by using the fifth-order COSY maps.
    The aperture sizes of the quadrupole magnets, the good-field regions in the dispersive plane, and the vertical interpole gap sizes in the non-dispersive plane of the dipole magnets were taken into account.
  }
  \label{fig:SS_HR_acceptances}
\end{figure}

\subsubsection{Momentum and angular acceptances}

The momentum and angular acceptances were evaluated by means of Monte-Carlo simulations using LISE$^{++}_{\mathit{cute}}$ with fifth-order COSY maps in the same way as for the HTBL as described in Section~\ref{subsection:AcceptancesHTBL}.
The results are shown in Fig~\ref{fig:SS_HR_acceptances}.
The angular acceptances show dependence on the momentum, and they are in excess of
$\lvert \varDelta x^\prime \rvert \lesssim \SI{50}{\mrad}$
and 
$\lvert \varDelta y^\prime \rvert \lesssim \SI{100}{\mrad}$
for the central momentum, and the momentum acceptance is
$\lvert \varDelta p/p \rvert \lesssim \SI{2.5}{\%}$
relative to the central momentum.

\begin{figure}[!t]
  \centering
  \includegraphics{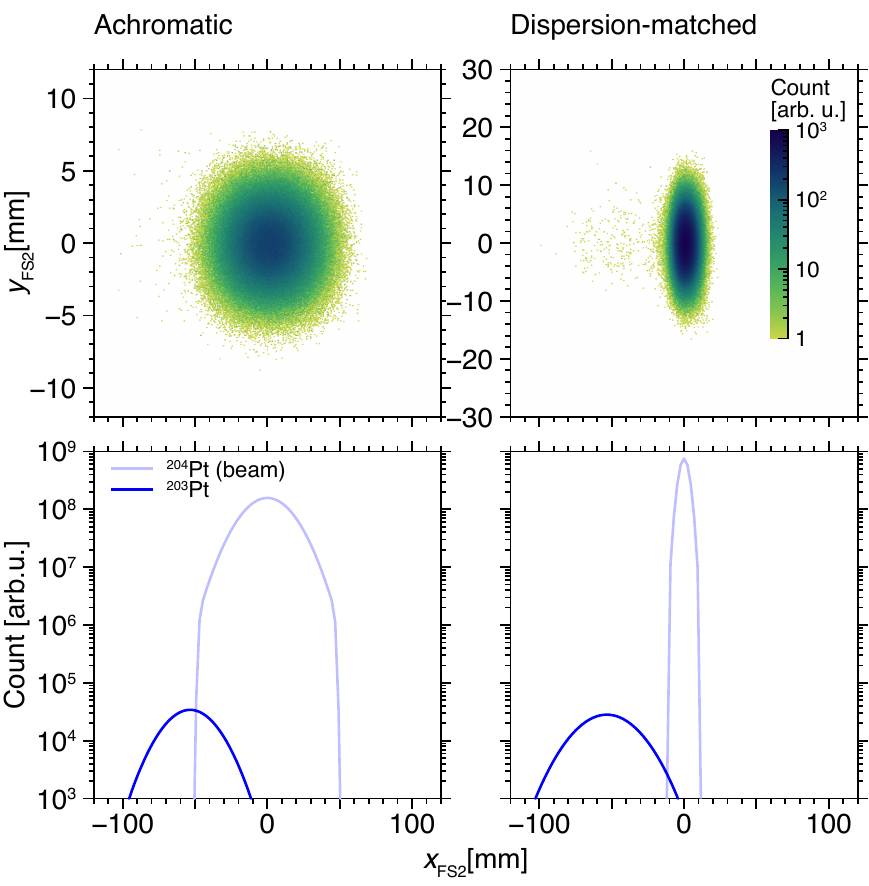}
  \caption{%
    Example of distributions at the spectrometer focal plane FS2 of the unreacted beam (${}^{204}\mathrm{Pt}$) and the reaction product (${}^{203}\mathrm{Pt}$) of the one-neutron-knockout reaction from ${}^{204}\mathrm{Pt}$ at \SI{100}{\MeV}/$u$ with a Be target with a thickness of \SI{100}{\mg/\cm^2}.
    The simulations were done in LISE$^{++}_{\mathit{cute}}$ with fifth-order COSY maps.
    The top panels show the $xy$ images from Monte-Carlo simulations using fifth-order maps and the bottom panels show one-dimensional distributions in first order generated in LISE$^{++}_{\mathit{cute}}$.
    The left and right panels are those with the HTBL in the achromatic mode and in the fully-dispersion-matched mode, respectively.
  }
  \label{fig:beam_images_FS2_comparison}
\end{figure}

\subsubsection{Blocking of unreacted beams}

Dispersion matching is useful, as discussed above, in blocking unreacted beams because the beam will be well localized at the achromatic focus to which the dispersion of the beamline is matched.
Figure~\ref{fig:beam_images_FS2_comparison} shows an example comparing the beam images at FS2 for the fully-dispersion-matched and achromatic beam transport for the one-neutron-knockout reaction from a $^{204}\mathrm{Pt}$ beam at \SI{100}{\MeV}/$u$ with a \SI{100}{\mg/\cm^2}-thick Be target simulated in LISE$^{++}_{\mathit{cute}}$ with transfer maps calculated in COSY.
As can be seen, the unreacted beam is more localized in the dispersive plane, intruding less into the region where the reaction products of interest, $^{203}\mathrm{Pt}$, are located when the beam is dispersion-matched than it is achromatically transported to the target.

Generally, dispersion matching enables a more efficient blocking of the beam, without having to obstruct the detection of the desired reaction products.
The unreacted-beam blocking must be optimized on an experiment-by-experiment basis, and in some cases partial dispersion matching may also be utilized instead to block the beam at the intermediate focal plane well before it reaches the final focal plane.

\subsubsection{ToF-\texorpdfstring{$B\rho$}{Brho} mass measurements}

\begin{figure}[!t]
  \centering
  \includegraphics{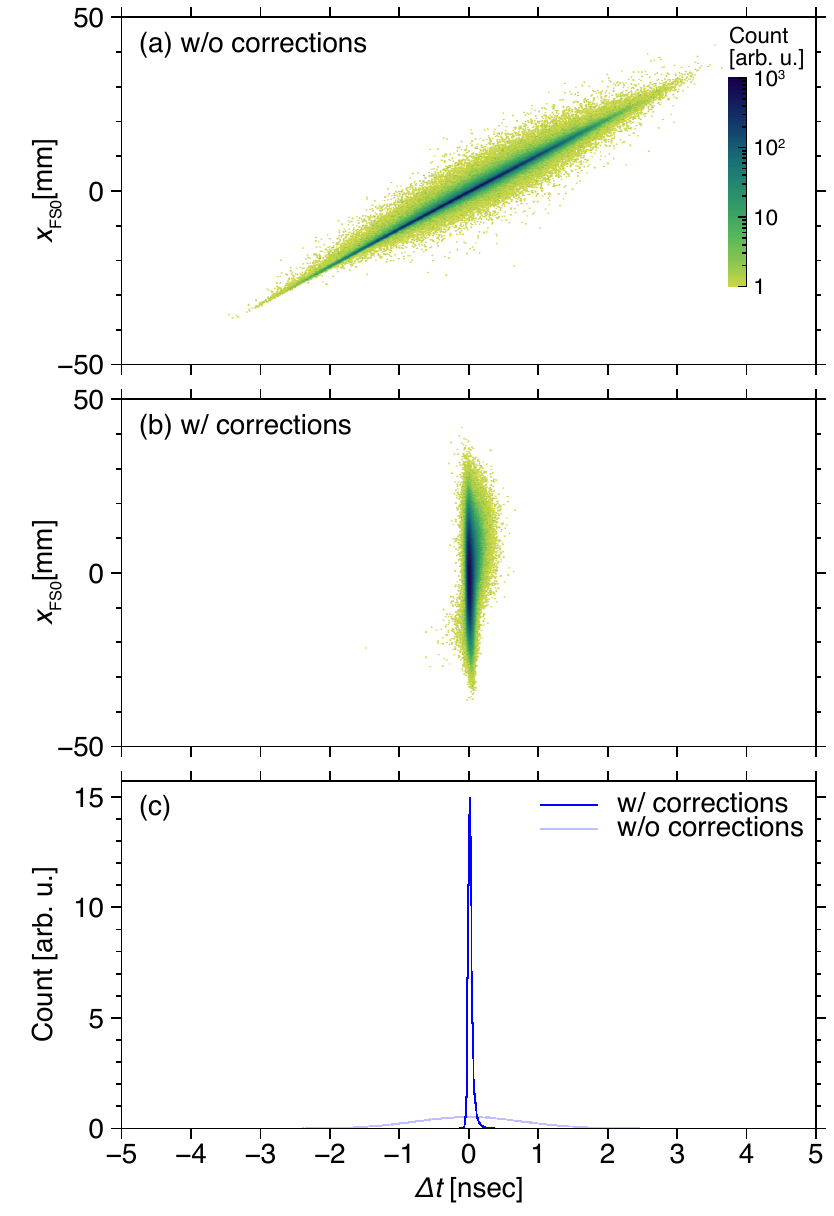}
  \caption{%
    (a) The correlations between $\varDelta t$ and $x_{\mathrm{FS0}}$ for ${}^{204}\mathrm{Pt}$ beam at \SI{4.0}{\Tm} transported through the reconfigured A1900 and the HRS.
    No corrections have been applied.
    (b) Same as (a) but corrections for $x_{\mathrm{FS0}}$ and $x^\prime_{\mathrm{FS2}}$ have been applied.
    (c) Projections of the two onto the horizontal axis.
    The $\varDelta t$ distributions without and with the corrections have widths of \SI{1800}{\ps} (FWHM) and \SI{50}{\ps} (FWHM), respectively.
  }
  \label{fig:ToFBrho_resolutions}
\end{figure}

The dispersion-matched beam transport is also utilized for mass-measurement experiments using the ToF-$B\rho$ technique, where the mass $m$ is deduced from the relation
\begin{equation*}
  m = \frac{q B\rho}{\gamma L / \mathrm{ToF}},
\end{equation*}
with $q$ being the charge of the particle and $L$ the flight-path length.
To create a long flight path, the entire HRS is combined with the second half of ARIS (the reconfigured A1900, which follows the Preseparator).
The length of the reconfigured A1900 is \SI{38.79}{\meter}, those of the HTBL and the Spectrometer Section are \SI{41.70}{\meter} and \SI{30.80}{\meter}, respectively, and the total length to be used for the ToF measurement is \SI{111.29}{\meter}.
The reconfigured A1900 is tuned achromatically.
The HTBL and the Spectrometer Section are fully dispersion-matched, rendering FS0 a dispersive focus, at which the magnetic rigidity is determined by measuring the dispersive position of the particle of interest, and FS2 an achromatic focus.
Due to the achromaticity, the beam is well localized at FS2, dedicated higher-resolution tracking detectors with small sensitive areas may be used there in lieu of the standard set of focal-plane detectors.
Two timing detectors will be used at both ends of the flight path to measure the ToF.

To evaluate the achievable mass resolution, Monte-Carlo simulations have been performed using fifth-order COSY maps.
A ${}^{204}\mathrm{Pt}$ beam at \SI{4.0}{\Tm} was used, and the position resolution of the detectors was assumed to be \SI{0.3}{\mm} (FWHM).
In COSY, the ToF is calculated as one of its coordinate variables having a dimension of length,
\begin{equation*}
  \ell = - (t-t_0) v_0 \frac{\gamma_0}{\gamma_0+1}
\end{equation*}
where $t_0$, $v_0$, and $\gamma_0$ are the ToF, the velocity, and the Lorentz factor of the reference particle, respectively, and
\begin{equation*}
  \varDelta t \equiv t - t_0 = - \frac{\ell}{v_0} \frac{\gamma_0+1}{\gamma_0}
\end{equation*}
is a measure of the deviation of the flight-path length from that of the reference particle which stems from the momentum and the angle deviations of each particle.
As can be seen in Fig.~\ref{fig:ToFBrho_resolutions}(a), $\varDelta t$ was strongly correlated with $x_{\mathrm{FS0}}$, which is a good measure of the momentum.
A phenomenological correction was applied such that the $x_{\mathrm{FS0}}$ dependence of $\varDelta t$ was eliminated and the locus became upright as in Fig.~\ref{fig:ToFBrho_resolutions}(b).
Similarly, $\varDelta t$ was also correlated with the angle $x^\prime_{\mathrm{FS2}}$, and a correction was applied to eliminate the correlation.
Figure~\ref{fig:ToFBrho_resolutions}(c) compares the $\varDelta t$ distributions before and after these corrections.
These results show that the momentum and angular corrections are effective in improving the ToF resolution which is deteriorated due to the flight-path-length uncertainty.

Once these corrections have been applied, the mass resolution is given by
\begin{equation}
  \frac{\varDelta m}{m} = \gamma ^2 \frac{\varDelta \mathrm{TOF}}{\mathrm{TOF}}.
\end{equation}
The right-hand side is the sum of the contribution from the path-length uncertainty as evaluated above, and the intrinsic timing resolution of the detectors, which is assumed here simply to be \SI{30}{\ps} (FWHM) based upon Refs.~\cite{NIMA.974.164199,NIMA.1027.166050}, albeit there is a charge dependence.
With this mass resolution $\varDelta m/m$, the statistical uncertainty in the mass determination $\delta m/m$, i.e.,~the determination of the centroid of the peak of the ToF spectrum from which the mass is deduced, is given with the peak counts $N$ by
\begin{equation}
  \frac{\delta m}{m} = \frac{1}{\sqrt{N}} \frac{\varDelta m}{m}.
\end{equation}
With $10^3$ peak counts, the achievable mass resolution $\delta m$ at mass $A = 50$, \num{100}, and \num{170} will be about \num{120}, \num{240}, and \SI{410}{\keV}, respectively.

\subsection{Neutron-invariant-mass mode}

\subsubsection{Ion-optical design}

The neutron-invariant-mass mode has been primarily designed for experiments that require detection of neutrons emitted at forward angles through the interpole gap of DS1, while providing a large momentum acceptance of beam-like fragments.
The reaction target is placed FS0$^\prime$ between QS2B and DS1, and FS1 is the final focal plane of this mode.
QS1A and QS2B become part of the beam transport to FS0$^\prime$, and QS5A, QS6A, and DS2 are not used.
DS1 is followed by two quadrupoles, QS3B and QS4B, which extend the flight path of charged particles while shaping their envelope, enabling a high mass resolving power without the need of focal-plane detectors with too large of the sensitive areas.
The layout allows the neutron-detector array MoNA-LISA to be installed at a distance of up to \SI{15}{\meter} from the target at FS0$^\prime$.
The location of QS3B has been determined such that it does not obstruct the neutrons emitted at forward angles.
The opening angles for the neutrons from FS0$^\prime$ through DS1 are about \SI{10}{\degree} horizontally and \SI{4.5}{\degree} vertically.
Note that ion-optical solutions all the way to FS2 come with reduced acceptances compared to utilizing FS1 as the final focal plane.
Also note that the aperture sizes of these quadrupoles are determined to achieve a \SI{10}{\msr} solid-angle acceptance in this mode;
these aperture sizes are not fully utilized in the high-resolution mode.
The moving capability of QS1A and QS2B would facilitate the optimization of the space around FS0$^\prime$ needed for installing beam-tracking detectors and other auxiliary detectors such as GRETA.

The first-order transfer-map elements are listed in Table~\ref{table:HRmatrixElements}, and the first-order ion-optical plots are shown in Fig.~\ref{fig:SS_envelopes_MEs}.
The two quadrupoles provide two degrees of freedom, which are used to achieve $(x \vert x^\prime) = (y \vert y) = 0$ at FS1.
A parallel-to-point imaging [$(y \vert y) = 0$] with a large value of $(y \vert y^\prime) = 2.49 \, \mathrm{m}/\mathrm{rad}$ is advantageous for achieving high accuracy in the reconstruction of the vertical component of the scattering angle.

\subsubsection{Momentum and angular resolutions}

\begin{figure}[!t]
  \centering
  \includegraphics{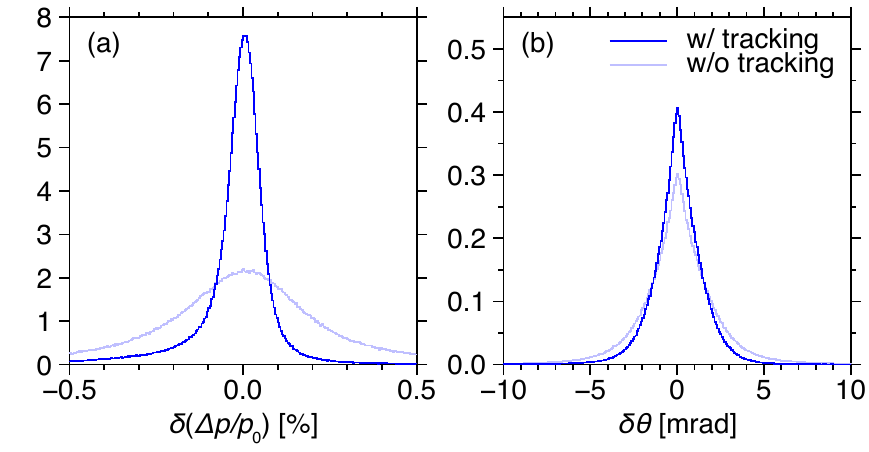}
  \caption{%
    Momentum and angular resolutions in the neutron-invariant-mass mode.
    The resolutions without and with beam tracking at FS0$^\prime$ are compared.
  }
  \label{fig:SS_NIM_resolutions}
\end{figure}

The momentum and angular resolutions were evaluated by means of Monte-Carlo simulations, in the same way as was done for the high-resolution mode.
The initial beam-spot size at FS0$^\prime$ was \SI{7.2}{\mm} ($x$) $\times$ \SI{20.5}{\mm} ($y$) in FWHM, as estimated for a ${}^{40}\mathrm{Mg}$ beam delivered through the HTBL and the two quadrupoles, QS1A and QS2B.
Note that the beam-spot size is large because of the large position magnifications of this beam transport which stem from the extended drift length preceding FS0$^\prime$.
This large beam-spot size deteriorates the accuracy of the momentum and angular reconstruction using an inverse map $\mathcal{S}$ as mentioned above.
Therefore, in addition to the reconstruction done with the inverse map $\mathcal{S}^{-1}$ from FS1 to FS0$^\prime$ as in Eq.~\eqref{eq:inverseMapping}, it was also done with a modified inverse map $\tilde{\mathcal{S}}^{-1}$ which takes into account, in addition to the positions and the angles at FS1, the dispersive position of the incoming beam particle at FS0$^\prime$ obtained from beam tracking, i.e.
\begin{equation} \label{eq:modInverseMapping}
  \begin{pmatrix}
    a _{\mathrm{FS0}^\prime} \\
    y _{\mathrm{FS0}^\prime} \\
    b _{\mathrm{FS0}^\prime} \\
    d _{\mathrm{FS0}^\prime}
  \end{pmatrix} _{\mathrm{recon}}
  = \tilde{\mathcal{S}} ^{-1}
  \begin{pmatrix}
    x _{\mathrm{FS1}} \\
    a _{\mathrm{FS1}} \\
    y _{\mathrm{FS1}} \\
    b _{\mathrm{FS1}} \\
    x _{\mathrm{FS0}^\prime}
  \end{pmatrix} _{\mathrm{meas}},
\end{equation}
in COSY notation.
Figure~\ref{fig:SS_NIM_resolutions} compares the momentum and angular resolutions calculated as the difference between the initial and reconstructed momenta and angles obtained via these two methods of reconstruction, i.e.~without and with $x _{\mathrm{FS0}^\prime}$ obtained from beam tracking.
The angular resolution was \SI{3.4}{\mrad} (FWHM) and \SI{2.8}{\mrad} (FWHM), and the momentum resolution was $1/\num{270} = 0.4\%$ (FWHM) and $1/\num{1300} = 0.08\%$ (FWHM) without and with beam tracking, respectively.
This deteriorated momentum resolution without beam tracking is due to the large beam-spot size in the dispersive plane ($x _{\mathrm{FS0}^\prime}$), which violates the assumption in the regular inverse mapping that the particle is emitted from a point source ($x _{\mathrm{FS0}^\prime} = \SI{0}{\mm}$) in the object plane.
The momentum resolution is recovered by including the dispersive position of the incoming beam particle in the modified reconstruction [Eq.~\eqref{eq:modInverseMapping}].

\begin{figure}[!t]
  \centering
  \includegraphics{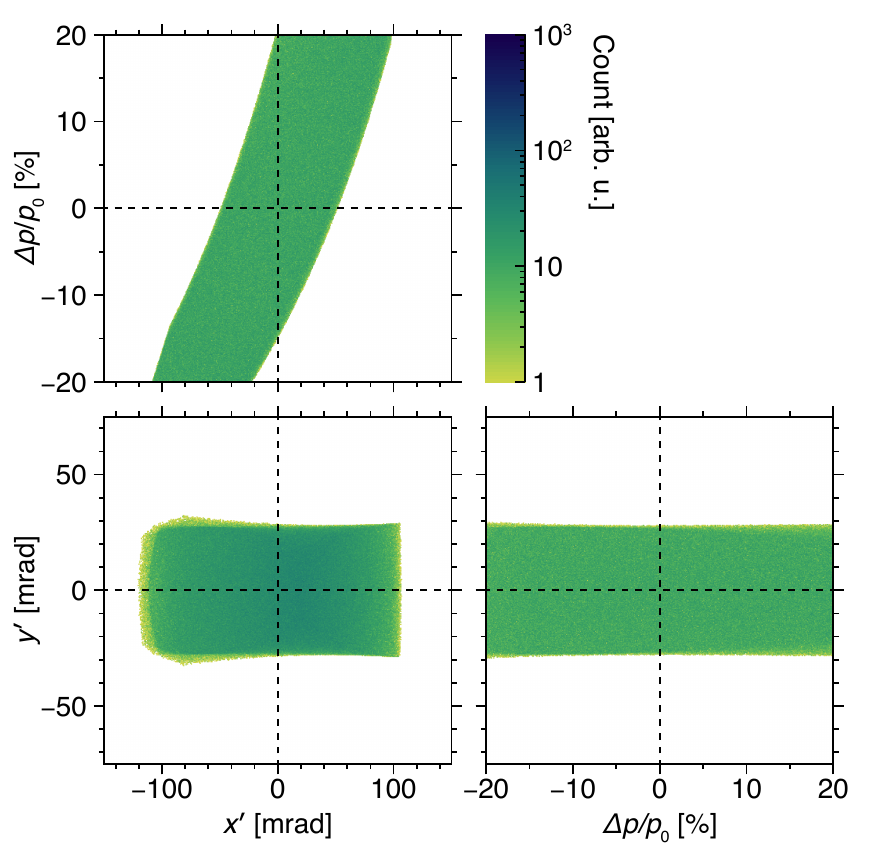}
  \caption{%
    Momentum and angular acceptances for the neutron-invariant-mass mode of the Spectrometer Section as simulated in LISE$^{++}_{\mathit{cute}}$ by using the fifth-order COSY maps.
    The aperture sizes of the quadrupole magnets, the good-field regions in the dispersive plane, and the vertical interpole gap sizes in the non-dispersive plane of the dipole magnets were taken into account.
  }
  \label{fig:SS_NIM_acceptances}
\end{figure}

\subsubsection{Momentum and angular acceptances}

The momentum and angular acceptances were evaluated by means of Monte-Carlo simulations using LISE$^{++}_{\mathit{cute}}$ with fifth-order COSY maps in the same way as for the high-resolution mode. 
Figure~\ref{fig:SS_NIM_acceptances} shows the results of these simulations.
Note that the angular and momentum acceptances correlate with each other more strongly than those in the high-resolution mode, because the first ion-optical element is a dipole magnet in this mode.
The angular acceptances are
$\lvert \varDelta x^\prime \rvert \lesssim \SI{100}{\mrad}$
and 
$\lvert \varDelta y^\prime \rvert \lesssim \SI{36}{\mrad}$,
and the solid-angle coverage is about \SI{14}{\msr}, satisfying the design objectives.
The momentum acceptance is significantly larger than the design objective of \SI{\pm5}{\%} around the center of the angular acceptance ($x^\prime \sim \SI{0}{\degree}$), and the centroid of the momentum coverage moves with the dispersive angle. 

\section{Magnets}
\label{section:magnets}

The magnets of the HRS have been designed based upon the ion-optical design described in Sections~\ref{section:ionOpticalDesignHTBL} and \ref{section:ionOpticalDesignSpectrometerSection}.
The realization of the ion-optical design requires magnets with large aperture sizes and relatively short lengths, therefore high field strengths must be driven by strong magnetomotive forces.
This has been the case for the S800~\cite{IEEETransApplSupercon.7.610} and the A1900~\cite{AdvCryoEng.43.245,NIMB.204.90} at NSCL, for which iron-dominated magnets driven by superconducting coils were implemented and demonstrated good performance.
Based upon this success, magnets of this type will be used for the HRS.
The designs of the HTBL magnets are described in Section~\ref{subsection:HTBLMagnets} and those of the Spectrometer Section magnets are described in Section~\ref{subsection:SpectrometerSectionMagnets}.

The ion-optical design and the magnet design have matured interactively, one informing the other in an iterative process to develop a consistent design.
In the COSY calculations presented above, use was already made of magnetic-field profiles obtained from calculations using the magnetostatic design described below in a three-dimensional finite element analysis (FEA) simulation software ANSYS Maxwell~\cite{Maxwell}, except for the HTBL quadrupole triplets, for which measured field-map data were available for various excitations covering the operational range from existing triplets of the same design~\cite{NIMB.376.150}.
The modeling of magnetic-field profiles for use in ion-optical calculations is described in Section~\ref{subsection:MagnetsInIonOpticalCalculations}.

\subsection{HTBL}
\label{subsection:HTBLMagnets}

\subsubsection{Dipole magnets}

\begin{figure}[!t]
  \centering
  \includegraphics{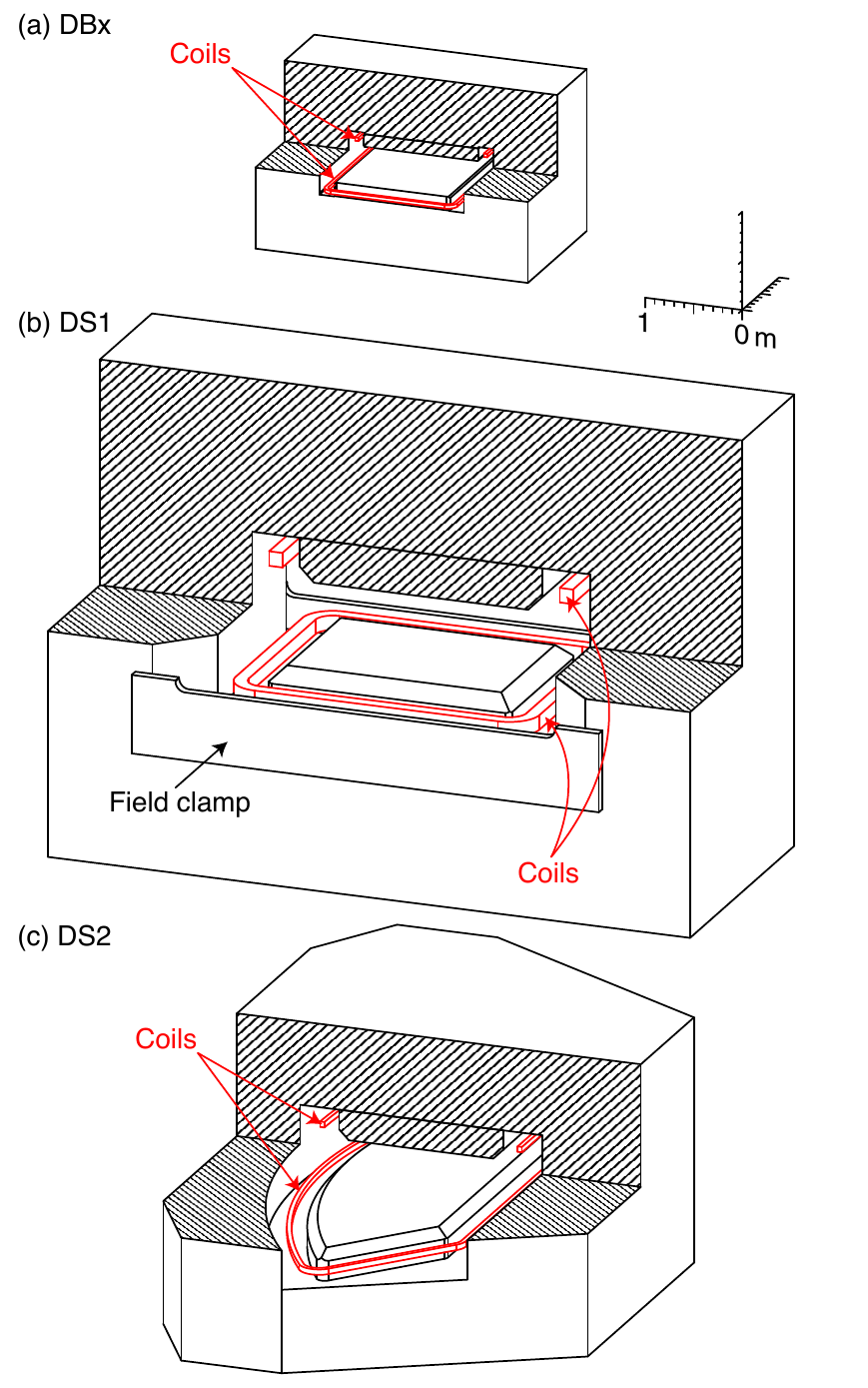}
  \caption{%
    Magnetostatic model of (a) DBx, (b) DS1, and (c) DS2.
    One quadrant is cut out to reveal the shapes of the iron yoke and the coils.
  }
  \label{fig:DBx_DS1_DS2_magnetostatic_models}
\end{figure}

The HTBL has four identical dipole magnets, DB1 through DB4. Their specifications are listed in Table~\ref{table:HTBLdipoles}.
The HTBL dipole design is similar to an existing rectangular dipole magnet used in the downstream of ARIS.
The iron and coil design of this dipole is shown in Fig.~\ref{fig:DBx_DS1_DS2_magnetostatic_models}(a).
The dimensions and the field strengths have been increased to accommodate the higher maximum magnetic rigidity of the HTBL.

\begin{table}[!t]
  \caption{%
    Specifications the dipole magnets of the HTBL.
    These are rectangular magnets, and the pole-face rotations cause horizontal focusing.
  }
  \centering
  \begin{tabularx}{\linewidth}{lXD{.}{.}{-1}}
    \toprule 
    Quantity                    &       & 4         \\
    Bending radius              & [m]   & 4.10      \\
    Maximum magnetic rigidity   & [Tm]  & 8.0       \\
    Maximum field               & [T]   & 2.0       \\
    Bending angle               & [deg] & 22.5      \\
    Arc length for central ray  & [m]   & 1.61      \\
    Vertical interpole gap size & [m]   & {\pm}0.05 \\
    Good-field-region width     & [m]   & {\pm}0.10 \\
    Entrance pole-face rotation & [deg] & 11.25     \\
    Exit pole-face rotation     & [deg] & 11.25     \\
    \bottomrule
  \end{tabularx}
  \label{table:HTBLdipoles}
\end{table}

\subsubsection{Quadrupole magnets}

\begin{figure}[!t]
  \centering
  \includegraphics{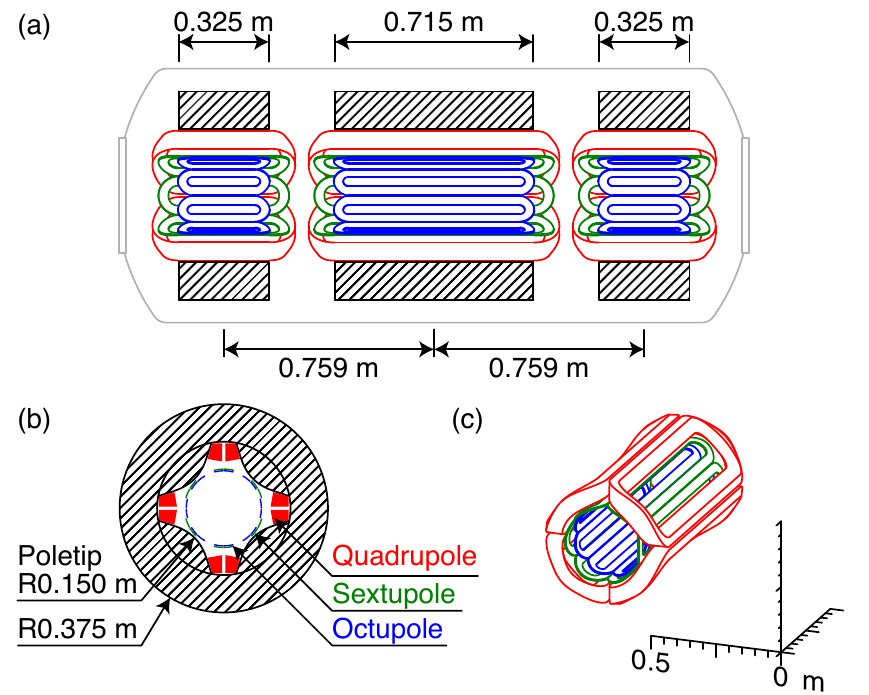}
  \caption{%
    (a) Cross-sectional view of the TBx quadrupole triplet as seen from the side.
    (b) Cross-sectional view as seen from the front. Those of both QB and QC quadrupoles are identical.
    (c) Magnetostatic model of the QC quadrupole. Only the coils are shown for better visibility.
  }
  \label{fig:NSCL_T7_triplet}
\end{figure}

The HTBL has eight identical quadrupole triplets, TB1 through TB8.
The design of one of three existing A1900 triplet quadrupole types~\cite{AdvCryoEng.43.245}, a picture of which is shown in Fig.~\ref{fig:NSCL_T7_triplet}, is chosen for the HTBL.
Since the A1900 has been reconfigured to be part of ARIS, the choice of this triplet design is appropriate for the HTBL so that matching acceptances can be achieved.
The configuration of this quadrupole triplet is QB-QC-QB in terms of the quadrupole names used in Ref.~\cite{AdvCryoEng.43.245}.
The specifications of these short QB and long QC quadrupoles are listed in Table~\ref{table:HTBLquads}.
Both quadrupole types have superimposed sextupole and octupole corrector coils.

\begin{table}[!t]
  \caption{%
    Specifications for the quadrupole magnets of the HTBL.
    The quadrupoles are packaged as QB-QC-QB triplets.
  }
  \centering
  \begin{tabularx}{\linewidth}{lX*{2}{D{.}{.}{-1}}}
    \toprule
    & & \multicolumn{1}{X}{\centering QB} & \multicolumn{1}{X}{\centering QC} \\
    \midrule
    Quantity                       &           & 16   & 8 \\
    Eff.~field length (nominal)    & [m]       & 0.40 & 0.79 \\
    Pole-tip radius                & [m]       & 0.15 & 0.15 \\
    Warm-bore radius               & [m]       & 0.10 & 0.10 \\
    Quadrupole max.~field gradient & [T/m]     & 17.1 & 18.2 \\
    Sextupole max.~field gradient  & [T/m$^2$] & 11   & 11 \\
    Octupole max.~field gradient   & [T/m$^3$] & 55   & 55 \\
    \bottomrule
  \end{tabularx}
  \label{table:HTBLquads}
\end{table}

\subsection{Spectrometer Section}
\label{subsection:SpectrometerSectionMagnets}

\begin{table}[!t]
 \caption{%
    Specifications for the dipole magnets of the Spectrometer Section.
    DS1 is a rectangular magnet, and the exit pole-face rotation causes horizontal defocusing.
    DS2 is a sector magnet, and both entrance and exit pole-face rotations cause horizontal focusing.
  }
  \centering
  \begin{tabularx}{\linewidth}{lX*{2}{D{.}{.}{-1}}}
    \toprule
    & & \multicolumn{1}{X}{\centering DS1} & \multicolumn{1}{X}{\centering DS2} \\
    \midrule
    Bending radius              & [m]   & 3.2       & 4.0       \\
    Maximum rigidity            & [Tm]  & 8.0       & 8.0       \\
    Maximum field               & [T]   & 2.5       & 2.0       \\
    Bending angle               & [deg] & -35       & 60        \\
    Arc length for central ray  & [m]   & 1.95      & 4.19      \\
    Vertical interpole gap size & [m]   & {\pm}0.30 & {\pm}0.10 \\
    Good-field-region width     & [m]   & {\pm}0.30 & {\pm}0.50 \\
    Entrance pole-face rotation & [deg] & 0         & -20       \\
    Exit pole-face rotation     & [deg] & 35        & -20       \\
    \bottomrule
  \end{tabularx}
  \label{table:SpectrometerSectionDipoles}
\end{table}

\subsubsection{Dipole magnets}

The Spectrometer Section has two dipole magnets of distinct designs, DS1 and DS2. 
Their specifications are listed in Table~\ref{table:SpectrometerSectionDipoles}.
The iron and coil designs of these dipoles are shown in Fig.~\ref{fig:DBx_DS1_DS2_magnetostatic_models}(b) and (c).

DS1 is a rectangular magnet.
It is placed such that the entrance pole face is perpendicular to the beam, and the exit pole-face rotation causes horizontal defocusing.
The vertical interpole gap size of \SI{60}{\cm} is determined so that it can accommodate the S{\textpi}RIT-TPC inside.
Given the high maximum field strength and considering that auxiliary detectors will be placed in close proximity of this dipole for some experiments in the neutron-invariant-mass mode, field clamps with a thickness of \SI{10}{\cm} are added to reduce the range of the fringe fields.

DS2 is a symmetric sector dipole with the pole-face rotations of \SI{-20}{\degree} both at the entrance and the exit which cause horizontal focusing.
Like the A1900 dipole magnets~\cite{AdvCryoEng.43.245}, the inner side of the coil is made straight to eliminate the negative curvature, which would make a coil winding more difficult and require a significant support structure to keep the coil in place.

\subsubsection{Quadrupole magnets}

\begin{figure}[!t]
  \centering
  \includegraphics{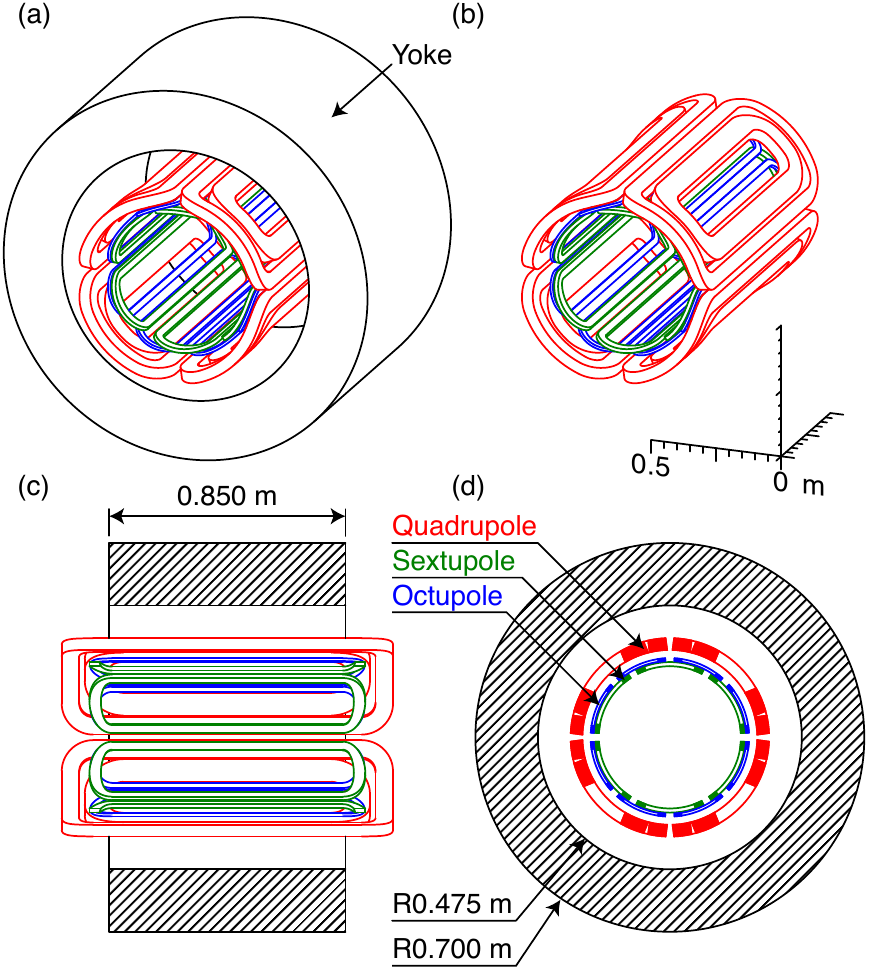}
  \caption{%
    (a) Magnetostatic model of the QSA quadrupole.
    (b) Only the coils are shown for better visibility.
    (c) Cross-sectional view as seen from the side.
    (d) Cross-sectional view as seen from the front.
  }
  \label{fig:QSA_magnetostatic_model}
\end{figure}

\begin{table}[!t]
  \caption{Specifications for the quadrupole magnets of the Spectrometer Section.}
  \centering
  \begin{tabularx}{\linewidth}{lX*{2}{D{.}{.}{-1}}}
    \toprule
    & & \multicolumn{1}{X}{\centering A} & \multicolumn{1}{X}{\centering B} \\
    \midrule
    Quantity                       &           & 3    & 3    \\
    Effective field length         & [m]       & 1.0  & 1.0  \\
    Warm-bore radius               & [m]       & 0.20 & 0.40 \\
    Quadrupole max.~field gradient & [T/m]     & 8.1  & 5.6  \\
    Sextupole max.~field gradient  & [T/m$^2$] & 5.0  & 2.6  \\
    Octupole max.~field gradient   & [T/m$^3$] & 9.0  & 2.8  \\
    \bottomrule
  \end{tabularx}
  \label{table:SpectrometerSectionQuads}
\end{table}

The Spectrometer Section has six quadrupole magnets of two different types.  
All the quadrupoles have superimposed sextupole and octupole corrector coils.
Their specifications are listed in Table~\ref{table:SpectrometerSectionQuads}.
Figure~\ref{fig:QSA_magnetostatic_model} shows the iron and coil designs of one of the two types (QSA).
The other type (QSB) is similar with different dimensions.

These quadrupoles are driven by four double-racetrack superconducting coils wrapping the surface of a cylinder.
The coils are contained by collars within the helium vessel and an iron yoke at room temperature. 
These magnets are modeled following the Cosine Two Theta quadrupole design of the Jefferson Lab Super High Momentum Spectrometer (SHMS)~\cite{IEEETransApplSupercon.18.415,IEEETransApplSupercon.19.1294}.
This design has the advantage of cylindrical vessels, minimized cold mass, and gives a compact cryostat.
The iron yoke is in the form of a simple tube, and owing to the absence of pole tips, the field is shaped dominantly by the coils.
This makes the field gradient approximately proportional to the excitation current throughout the operating range as shown in Fig.~\ref{fig:QSA_b20}(b).
The sextupole and octupole corrector coils are packaged similarly to the HTBL quadrupoles and the other warm-iron quadrupoles of ARIS.

\subsection{Modeling of magnetic-field profiles for ion-optical calculations}
\label{subsection:MagnetsInIonOpticalCalculations}

The magnetic-field profiles that were used in the COSY calculations presented above were obtained in magnetostatic calculations performed in ANSYS Maxwell, except for the HTBL quadrupole triplets, for which field profiles obtained from field-map data measured using an existing A1900 quadrupole triplet of the same design~\cite{NIMB.376.150} were used.
In the following, the modeling of the magnetic-field profiles of the dipole fields and that of the multipole (quadrupole, sextupole, and octupole) fields are described in Sections~\ref{subsubsection:MagnetsInIonOpticalCalculationsDipoles} and \ref{subsubsection:MagnetsInIonOpticalCalculationsQuadrupoles}, respectively.

\subsubsection{Dipoles}
\label{subsubsection:MagnetsInIonOpticalCalculationsDipoles}

For the description of the dipole fields, the MSS model in COSY~\cite{COSYBeamMan91,NIMB.376.150} was used.
In this model, the field in the median plane is expressed as
\begin{equation*}
  B_y (y = 0) = \frac{\chi_M}{\rho_0} \cdot E_{\mathrm{ent}} \cdot E_{\mathrm{ex}} \cdot W(\rho,\phi),
\end{equation*}
where $\chi_M$ is the magnetic rigidity of the reference particle and $\rho_0$ is the radius of the reference trajectory.
$E_{\mathrm{ent}}$ and $E_{\mathrm{ex}}$ are the Enge functions that describe the fringe fields at the entrance and the exit, respectively, with the Enge function $E (\xi)$ being in the form
\begin{equation*}
  E(\xi) = \dfrac{1}{1 + \exp \left[ \displaystyle \sum _{k = 1} ^N a _k \left(\frac{\xi}{D}\right) ^{k-1} \right]},
\end{equation*}
where $\xi$ is the distance perpendicular to the effective field boundary, $D$ is the full aperture, and COSY employs $N = 6$ coefficients.
The dipole magnets of the HRS have large aperture sizes relative to their lengths, which results in little to no flat field regions especially at high excitation currents.
Therefore, a convoluted form of the Enge functions, $E_{\mathrm{ent}} \cdot E_{\mathrm{ex}}$, was used. 
A two-dimensional function $W (\rho,\phi)$ describes the radial ($\rho$) and azimuthal ($\phi$) dependences as
\begin{equation*}
  W (\rho,\phi) = \sum_{i = 0}^k \sum_{j = 0}^k w _{ij} (\rho - \rho_0) ^i \left(\phi - \frac{\phi_0}{2}\right) ^j,
\end{equation*}
where the coefficients $w_{ij}$ were obtained by fitting the field profiles.
While the standard COSY accommodates coefficients up to $k = 2$, in the enhanced version used for the present study these have been extended to $k = 6$ to better describe the field profiles~\cite{NIMB.376.150}.
From these parameters, the field in the entire beam region is calculated internally by means of the high-order out-of-plane expansion such that it satisfies Maxwell's equations~\cite{IJMPA.25.1807}.

\begin{figure}[!t]
  \centering
  \includegraphics{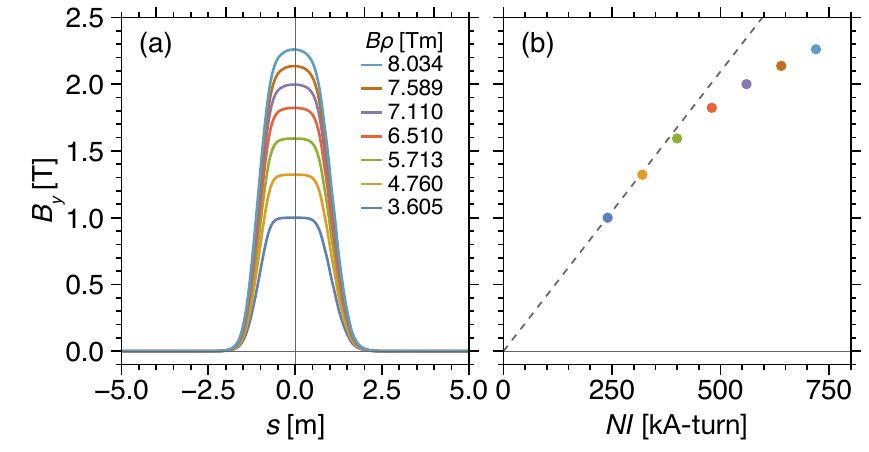}
  \caption{%
    (a) DS1 field profiles along the central trajectory for various excitation currents covering the operational magnetic-rigidity range.
    (b) Maximum field strength at each excitation current.
    The current value on the horizontal axis is for only one of the two coils.
    The dashed line represents the calculated field strength on the assumption that iron has infinite permeability (no effects of saturation).
  }
  \label{fig:DS1_field_profiles}
 
  \includegraphics{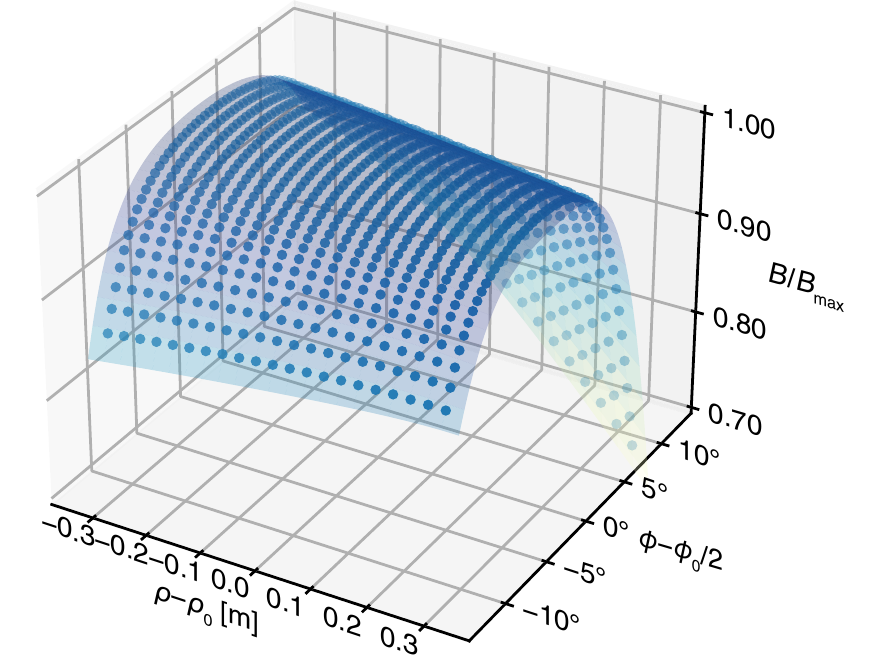}
  \caption{%
    Radial and azimuthal dependences of the DS1 field profile at its maximum excitation current.
    Sample points and the fitted $W(\rho,\phi)$ function shown as a surface.
  }
  \label{fig:DS1_2D_field_shape}
\end{figure}

Magnetostatic calculations in ANSYS Maxwell were performed at various excitation currents so as to cover the operational magnetic-rigidity range.
The field profile of DS1 along the central trajectory is shown in Fig.~\ref{fig:DS1_field_profiles}(a) for various excitation currents together with the excitation curve in Fig.~\ref{fig:DS1_field_profiles}(b).
An example of the radial and azimuthal dependences of the DS1 field profile is shown in Fig.~\ref{fig:DS1_2D_field_shape} at its maximum excitation current.
These parameters and the excitation dependence thereof were included in the COSY calculations.

\subsubsection{Multipoles}
\label{subsubsection:MagnetsInIonOpticalCalculationsQuadrupoles}

\begin{figure}[!t]
  \centering
  \includegraphics{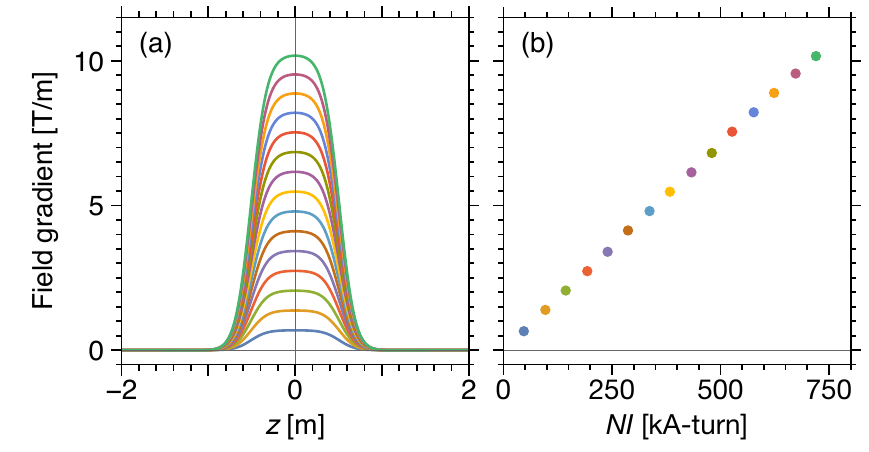}
  \caption{%
    (a) QSA field gradients from $b_{2,0}(z)$ for various excitation currents covering the operational magnetic-rigidity range.
    (b) Maximum field gradient at each excitation current.
    The current value on the horizontal axis is the sum of the two coils in one quadrant.
  }
  \label{fig:QSA_b20}

  \includegraphics{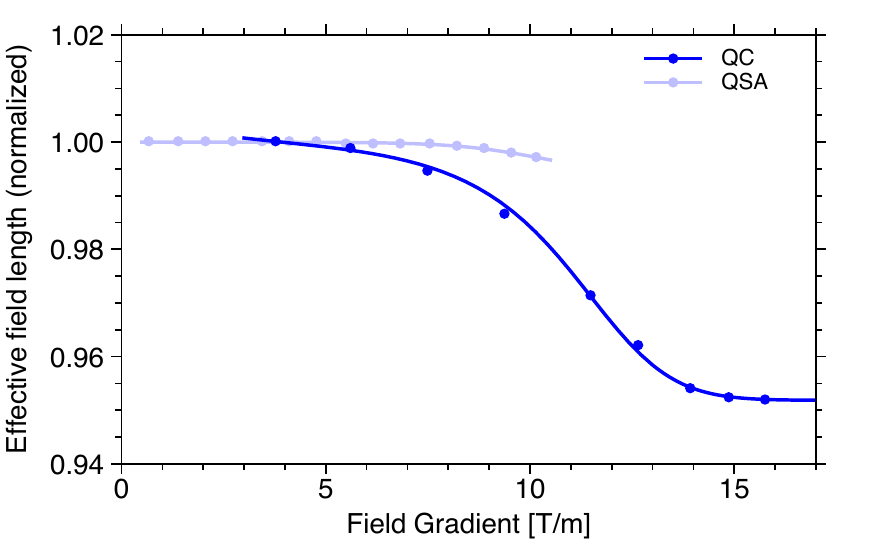}
  \caption{%
    Effective field lengths of QC and QSA for various excitation currents.
    The data points are normalized such that the maximum effective field lengths is $1$ for each quadrupole.
    The curves represent the fitted analytical equations which were used in the COSY calculations to include the excitation dependence of the field lengths.
  }
  \label{fig:QC_QSA_effective_field_lengths}
\end{figure}

The multipole field has a straight beam axis and a certain rotational symmetry.
The analysis of the calculated or measured field has been done according to the method described in Ref.~\cite{NIMB.317.798}, in which the field is expressed in a form of the Fourier--Bessel series which, in the cylindrical coordinate system, $(\rho,\phi,z)$, is written as
\begin{multline*}
  \varPhi (\rho,\phi,z)
  = \sum_{n=1}^\infty \sum_{m=0}^\infty
  \left[
    \frac{b_{n,m}(z) \rho _0}{n + 2 m}
    \left(
      \frac{\rho}{\rho _0}
    \right) ^{n + 2 m}
    \sin n \phi 
  \right. \\
  +
  \left.
    \frac{a_{n,m}(z) \rho _0}{n + 2 m}
    \left(
      \frac{\rho}{\rho _0}
    \right) ^{n + 2 m}
    \cos n \phi
  \right],
\end{multline*}
where the second term (the skew term) vanishes for a multipole with pure midplane symmetry.
The $z$-dependent functions $b_{n,m}(z)$ are a generalized gradient, and the pure quadrupole, sextupole, and octupole components correspond to $b_{2,0}(z)$, $b_{3,0}(z)$, $b_{4,0}(z)$, respectively.
As an example, $b_{2,0}(z)$ for the QSA quadrupole in the form of the field gradient is shown in Fig.~\ref{fig:QSA_b20}(a) for various excitation currents.

The analysis requires field data obtained from the magnetostatic calculations on the surface of a cylinder with equally spaced azimuthal angles.
For the present analysis, the radius of the cylinder was set at \SI{80}{\%} of the bore aperture, and the angle step $\pi/64 = \SI{2.8125}{\degree}$.

From the $b_{n,0}(z)$ functions ($n=2$, $3$, or $4$) the effective field length
\begin{equation*}
  L _{\mathrm{eff}} = \frac{1}{b_{n,0} (z = 0)} \int \mathrm{d} z \, b_{n,0}(z)
\end{equation*}
and the Enge coefficients were extracted.
Due to the relatively short length compared to the aperture size, a convoluted form of the Enge functions~\cite{NIMB.317.798} was used also for the multipoles, where $b_{n,0}(z)$ was expressed as
\begin{equation*}
  b _{n,0}(z) = B_c \cdot E _{\mathrm{ent}} (-[z+L_{\mathrm{eff}}/2]) \cdot E _{\mathrm{ex}} (+[z-L_{\mathrm{eff}}/2]).
\end{equation*}
Because the shape of $b_{n,0}(z)$ changes as a function of excitation current, the magnetostatic calculations were performed at various excitation currents so as to cover the operational magnetic-rigidity range.
The effective field lengths decrease as the field strengths increase, as can be seen in Fig.~\ref{fig:QC_QSA_effective_field_lengths} which shows the effective field lengths of QC and QSA as a function of the field gradient.
Note that the variation of the length is smaller for the current-dominant QSA than for the iron-dominant QC.
Such variations of the effective field lengths and those of the Enge coefficients were taken into account in the COSY calculations.

\section{Detectors}
\label{section:detectors}

The HRS will have beam-diagnostic equipment and detectors that characterize beam properties to facilitate beam tuning and provide event-by-event beam-tracking and particle-identification information.
The basic concepts and operational conditions are based upon the experience gained from NSCL operations~\cite{PhysScr.91.053003}, including the A1900, the S800~\cite{NIMA.422.291}, and the Sweeper Magnet.
In the following, the detector configurations are described in Section~\ref{subsection:detectorConfigurations} and the data-acquisition (DAQ) system in Section~\ref{subsection:DAQ}.

\begin{figure}[!t]
  \centering
  \includegraphics{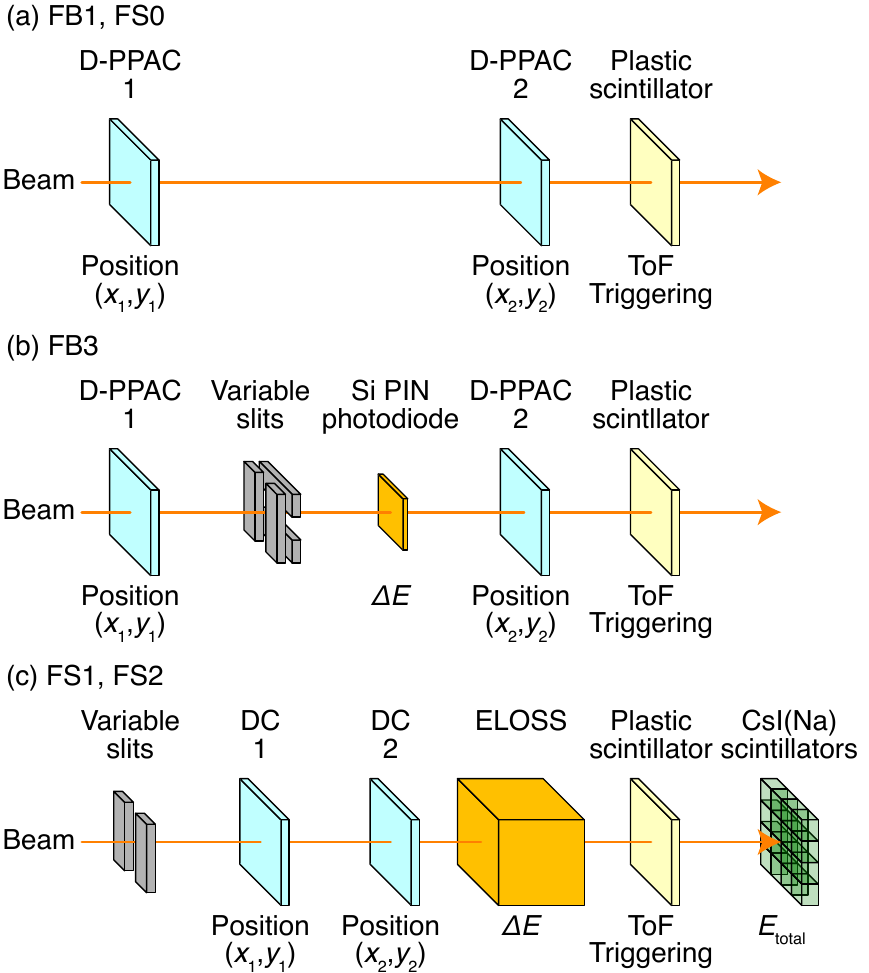}
  \caption{%
    Schematic layouts of the detector configurations at (a) FB1 and FS0, (b) FB3, and (c) FS1 and FS2. 
    Although not included, luminescent beam viewers will also be installed in all the HTBL focal planes.
    See text for details.
  }
  \label{fig:diagnostics_layouts}
\end{figure}

\subsection{Detector configurations and specifications}
\label{subsection:detectorConfigurations}

The schematic layout of the detector configuration at each focal plane of the HRS is shown in Fig.~\ref{fig:diagnostics_layouts}.

In the HTBL, pairs of two-dimensional position-sensitive detectors at the dispersive focal planes in the achromatic mode, FB1 and FB3, enable beam tracking at these focal planes, from which the angle and momentum of the incoming beam at the spectrometer target can be inferred.
Both FB1 and FB3 will be equipped with delay-line parallel-avalanche counters (D-PPACs) as well as a plastic scintillator as a timing detector.
The position and timing resolutions better than \SI{1}{\mm} (FWHM) and \SI{150}{\ps} (FWHM), respectively, should be achieved for these detectors.
In addition, FB3 will have an energy-loss ($\varDelta E$) detector, a silicon PIN diode, for particle identification, and a variable slit system for background rejection, which are necessary for the operation of the RFFS installed at FB2.
At FS0, the same set of detectors as FB1 will be made available and used when needed, but the detailed configuration may vary from one experiment to another;
for example, a micro-channel-plate (MCP) detector can be used here for precise position determination at high rates for ToF-$B\rho$ mass-measurement experiments, which require magnetic-rigidity determination at this dispersive focus.
All the focal planes will have luminescent beam viewers that provide visual feedback in beam tuning.
All of these will be made able to be remotely inserted in and retracted from the beam path on pneumatic or linear drives.

In the Spectrometer Section, the focal planes are equipped with detectors for trajectory reconstruction and particle identification of the reaction products.
The detectors at FS1 and FS2 consist of a pair of two-dimensional position-sensitive detectors, an energy-loss detector, a timing detector, and a total-kinetic-energy detector.
The position-sensitive detectors are based upon drift chambers (DCs) that are being newly constructed for the S800~\cite{JInst.15.P03025}, replacing the Cathode-Readout Drift Chamber (CRDC) that have been used previously~\cite{NIMA.422.291}.
This new DC design employs an advanced hybrid micro-pattern gaseous detector readout, consisting of a multi-layer Thick Gaseous Electron Multiplier (M-THGEM)~\cite{RevSciInstrum.88.013303} mounted on top of a position-sensitive MICROMEGAS~\cite{NIMA.376.29}.
A position resolution better than \SI{1}{\mm} (FWHM) should be able to be obtained.
The energy-loss detectors will be Energy-Loss Optical Scintillation System (ELOSS)~\cite{PANIC2021.3398}, the first version of which is being constructed for the S800.
They are filled with high-scintillation-yield gas, such as xenon, and intense, prompt scintillation light emitted along the track of a beam particle that traverse the sensitive volume will be read out by an array of photodetectors that surround the volume.
As mentioned above, an energy-loss resolution should be better than \SI{1.3}{\%} (FWHM) to achieve $4\sigma$ separation between isotopes with neighboring charge number up to $Z = 92$.
The timing detectors will be plastic scintillators having large active areas coupled to multiple photomultiplier tubes.
This design enables an optimal photon-collection efficiency and an expected time resolution of the order of \SI{150}{\ps} (FWHM).
Finally, the total-kinetic-energy ($E_{\mathrm{total}}$) detectors will consist of hodoscopes made of CsI(Na) scintillators, similar to those currently implemented at the S800~\cite{NIMA.652.668}.
In addition, variable slits will be included at both FS1 and FS2.
All of these at FS1 will need to be remotely inserted in and retracted from the beam path on pneumatic or linear drives, while those at FS2 can be stationary.

A rate capability in excess of \SI{1}{\mega\Hz} should be achieved for the timing detectors.
Beamline scintillators can be and actually have been routinely operated at rates in excess of 1 MHz at NSCL.
Note that, aside from scaler readout for counting statistics, timing is only stored for events that are selected by an external trigger, e.g., from the focal plane of the Spectrometer Section, whose rate will be much lower (maximally \SI{\sim10}{\kilo\Hz}).
The same is true for beamline D-PPACs, which can operate at a few hundred kHz, but as is the case for the timing detectors, data will be read out and stored only for events selected by an external trigger, as can be done with existing technology.

\subsection{Data-acquisition (DAQ) system}
\label{subsection:DAQ}

Signals from the detectors must be read out by an appropriate DAQ system so that they can be recorded and analyzed on an event-by-event basis and those from different focal planes can be correlated.
The existing DAQ system used at NSCL (NSCLDAQ)~\cite{NSCLDAQ} is capable of such distributed data acquisition and is suited for the DAQ system of the HRS.
The timestamp-based DAQ synchronization has been used in many of the S800 experiments.
Some examples for the DAQ synchronization include that with GRETINA (Gamma-Ray Energy Tracking In-beam Nuclear Array)~\cite{NIMA.847.187}, with LENDA (Low-Energy Neutron-Detector Array)~\cite{NIMA.815.1}, and with both GRETINA and LENDA~\cite{PhysRevLett.122.232701}.
These serve as a model for the DAQ integration of the HRS and the auxiliary detectors.
NSCLDAQ has been continuously improved and is adapted for new hardware in close coordination and collaboration with experimenters. 
Present developments focus on high-rate data streaming and parallel processing capabilities, which will also benefit HRS experiments. 

Any analysis framework, such as ROOT~\cite{NIMA.389.81} or NSCLSpecTcl~\cite{NSCLSpecTcl} can be attached to the data flow to perform online and offline analyses.
NSCLSpecTcl provides a variety of histogram and gate types, and its user interface aids in dynamic creation of histograms and application of gates.
Detailed analyses facilitated by such framework is crucial for the commissioning and operation of the HRS. 
Real-time data processing is important for immediate assessment of beam tune by examining correlations between phase-space parameters at different focal planes, which enables quantification of ion-optical transfer-map elements, as has been routinely done at the A1900 and the S800.
Such functionality is also useful to evaluate the quality of physics data.

\section{Summary and conclusion}

An ion-optical design for the HRS consisting of the spectrometer proper, the Spectrometer Section, and the preceding analysis beamline, the HTBL, has been described.
Detailed ion-optical simulations using the realistic magnetic-field profiles have been performed to examine various aspects of the beam transport through the HRS, and it has been demonstrated that the design meets the specifications.
The specifications of the magnet and detector systems have been stipulated.

\currentpdfbookmark{Declaration of competing interest}{ConflictOfInterest}
\section*{Declaration of competing interest}

The authors declare that they have no known competing financial interests or personal relationships that could have appeared to influence the work reported in this paper.

\currentpdfbookmark{Data availability}{AataAvailability}
\section*{Data availability}

Data will be made available on request.

\currentpdfbookmark{Acknowledgments}{Acknowledgements}
\section*{Acknowledgments}

The authors are grateful to Dr.~Hiromi Sato for his work on the initial magnetostatic and mechanical design of the magnets, and also to the members of the Magnet Design Review group for important suggestions.
The science specifications have been devised on the basis of frequent feedback from the members of the HRS Working Group.
This material is based upon work supported by the U.S. Department of Energy, Office of Science, Office of Nuclear Physics under grant Nos.~DE-SC0014554 and DE-SC0000661 (FRIB) and contract No.~DE-AC02-06CH11357 (ANL).

\appendix
\section{Dispersion matching}
\label{Appendix:DM}

The deduction of the changes in the dispersive angle and the momentum of the beam particle introduced by the reaction (scattering) that takes place in the target in first order when the beamline is dispersion-matched to the spectrometer is described.
Two cases are considered below, namely the case where the beamline is dispersion-matched to the full spectrometer (full dispersion matching), and the case where the beamline is dispersion-matched to only a part of the spectrometer (partial dispersion matching).

\subsection{Full dispersion matching}

Let the position, angle, and momentum $(x, x^\prime, \delta)$ in the dispersive plane at the object, the target, and the focal plane be $\vec{X}_{\mathrm{obj}}$, $\vec{X}_{\mathrm{tgt}}$, and $\vec{X}_{\mathrm{fp}}$, respectively.
Let the transfer matrices of the beamline and  the spectrometer be $\mathit{M} _{\mathrm{B}}$ and $\mathit{M} _{\mathrm{S}}$, respectively.
Let changes in the angle and momentum induced by the reaction (scattering) be $\vec{X}_{\mathrm{sc}} = (0, x^\prime _{\mathrm{sc}}, \delta _{\mathrm{sc}})$.
With these, the coordinates at the focal plane after the reaction are expressed as
\begin{equation} \label{eq:dm_X_S2}
  \vec{X}_{\mathrm{fp}} 
  = \mathit{M} _{\mathrm{S}} (\mathit{M} _{\mathrm{B}} \vec{X}_{\mathrm{obj}} +
  \vec{X}_{\mathrm{sc}}) ,
\end{equation}
or more explicitly,
\begin{multline*}
  \begin{pmatrix}
    x _{\mathrm{fp}} \\
    x^\prime _{\mathrm{fp}} \\
    \delta _{\mathrm{fp}}
  \end{pmatrix}
  = 
  \begin{pmatrix}
    (x \vert x) _{\mathrm{S}} & 0 & (x \vert \delta) _{\mathrm{S}} \\
    (x^\prime \vert x) _{\mathrm{S}} & (x^\prime \vert x^\prime) _{\mathrm{S}} & (x^\prime \vert \delta) _{\mathrm{S}} \\
    0 & 0 & 1
  \end{pmatrix} \\
  \left[
  \begin{pmatrix}
    (x \vert x) _{\mathrm{B}} & 0 & (x \vert \delta) _{\mathrm{B}} \\
    (x^\prime \vert x) _{\mathrm{B}} & (x^\prime \vert x^\prime) _{\mathrm{B}} & (x^\prime \vert \delta) _{\mathrm{B}} \\
    0 & 0 & 1
  \end{pmatrix}
  \begin{pmatrix}
    x _{\mathrm{obj}} \\
    x^\prime _{\mathrm{obj}} \\
    \delta _{\mathrm{obj}}
  \end{pmatrix}
  +
  \begin{pmatrix}
    0 \\
    x^\prime _{\mathrm{sc}} \\
    \delta _{\mathrm{sc}}
  \end{pmatrix}
  \right] ,
\end{multline*}
where the point-to-point focus conditions
\begin{equation*}
  (x \vert x^\prime) _{\mathrm{B}} = (x \vert x^\prime) _{\mathrm{S}} = 0 .
\end{equation*}
were assumed. 
This gives
\begin{equation} \label{eq:dm_xS2_aS2_wodm}
  \begin{aligned}
  x _{\mathrm{fp}}
  =& \, (x \vert x) _{\mathrm{S}} (x \vert x) _{\mathrm{B}} x _{\mathrm{obj}}
  + [(x \vert x) _{\mathrm{S}} (x \vert \delta) _{\mathrm{B}} 
    + (x \vert \delta) _{\mathrm{S}} ] \delta _{\mathrm{obj}}
    + (x \vert \delta) _{\mathrm{S}} \delta _{\mathrm{sc}} , \\
  x^\prime _{\mathrm{fp}}
  =& \, [(x^\prime \vert x) _{\mathrm{S}} (x \vert x) _{\mathrm{B}} 
    + (x^\prime \vert x^\prime) _{\mathrm{S}} (x^\prime \vert x) _{\mathrm{B}}] x _{\mathrm{obj}} \\
  &+ (x^\prime \vert x^\prime) _{\mathrm{S}} (x^\prime \vert x^\prime) _{\mathrm{B}} x^\prime _{\mathrm{obj}} 
    + (x^\prime \vert x^\prime) _{\mathrm{S}} x^\prime _{\mathrm{sc}} \\
  &+ [(x^\prime \vert x) _{\mathrm{S}} (x \vert \delta) _{\mathrm{B}} 
    + (x^\prime \vert x^\prime) _{\mathrm{S}} (x^\prime \vert \delta) _{\mathrm{B}}
    + (x^\prime \vert \delta) _{\mathrm{S}} ] \delta _{\mathrm{obj}} \\
  &+ (x^\prime \vert \delta) _{\mathrm{S}} \delta _{\mathrm{sc}} .
  \end{aligned}
\end{equation}
The dispersion-matching conditions are that the coefficients of $x _{\mathrm{fp}}$ and $x^\prime _{\mathrm{fp}}$ with respect to $\delta _{\mathrm{obj}}$ vanish, namely, 
\begin{subequations} \label{eq:dm_condition}
  \begin{align}
  (x \vert x) _{\mathrm{S}} (x \vert \delta) _{\mathrm{B}} 
    + (x \vert \delta) _{\mathrm{S}} \label{eq:dm_condition_x}
  &= 0 ,\\
  (x^\prime \vert x) _{\mathrm{S}} (x \vert \delta) _{\mathrm{B}} 
    + (x^\prime \vert x^\prime) _{\mathrm{S}} (x^\prime \vert \delta) _{\mathrm{B}}
    + (x^\prime \vert \delta) _{\mathrm{S}} \label{eq:dm_condition_a}
  &= 0 ,
  \end{align}
\end{subequations}
with which Eqs.~\eqref{eq:dm_xS2_aS2_wodm} are simplified as
\begin{subequations} \label{eq:dm_xS2_aS2}
\begin{align}
  x _{\mathrm{fp}} 
    ={}& (x \vert x) _{\mathrm{S}} (x \vert x) _{\mathrm{B}} x _{\mathrm{obj}}
    + (x \vert \delta) _{\mathrm{S}} \delta _{\mathrm{sc}} , \label{eq:dm_xS2_aS2_x} \\
  x^\prime _{\mathrm{fp}} 
    ={}& [(x^\prime \vert x) _{\mathrm{S}} (x \vert x) _{\mathrm{B}} 
    + (x^\prime \vert x^\prime) _{\mathrm{S}} (x^\prime \vert x) _{\mathrm{B}}] x _{\mathrm{obj}} \nonumber \\
    &+ (x^\prime \vert x^\prime) _{\mathrm{S}} (x^\prime \vert x^\prime) _{\mathrm{B}} x^\prime _{\mathrm{obj}} 
    + (x^\prime \vert x^\prime) _{\mathrm{S}} x^\prime _{\mathrm{sc}}
    + (x^\prime \vert \delta) _{\mathrm{S}} \delta _{\mathrm{sc}}. \label{eq:dm_xS2_aS2_alpha}
\end{align}
\end{subequations}
The momentum change $\delta _{\mathrm{sc}}$ is obtained from Eq.~\eqref{eq:dm_xS2_aS2_x} as
\begin{align}
  \delta _{\mathrm{sc}}
    &= \frac{1}{(x \vert \delta) _{\mathrm{S}}} 
    [x _{\mathrm{fp}} - (x \vert x) _{\mathrm{S}} (x \vert x) _{\mathrm{B}} 
      x _{\mathrm{obj}}] \nonumber \\
    &= \frac{x _{\mathrm{fp}}}{(x \vert \delta) _{\mathrm{S}}} 
      + \frac{(x \vert x) _{\mathrm{B}}}{(x \vert \delta) _{\mathrm{B}}} x _{\mathrm{obj}} , \label{eq:dm_deltasc}
\end{align}
where Eq.~\eqref{eq:dm_condition_x} was used.
Note that $\delta _{\mathrm{sc}}$ does not depend on $\delta _{\mathrm{obj}}$ of the beam particle, i.e. $\delta _{\mathrm{sc}}$ can be deduced without having to measure $\delta _{\mathrm{obj}}$ as a consequence of dispersion matching.
As in the second term on the RHS in Eq.~\eqref{eq:dm_deltasc}, a measurement of the position $x _{\mathrm{obj}}$ is needed for the deduction of $\delta _{\mathrm{sc}}$.
However, because the beam-spot size at the object $x _{\mathrm{obj}}$ is usually small, the measurement thereof may be omitted.

The angle change $x^\prime _{\mathrm{sc}}$ is obtained from Eq.~\eqref{eq:dm_xS2_aS2_alpha} as
\begin{multline*}
  x^\prime _{\mathrm{sc}}
    = \frac{1}{(x^\prime \vert x^\prime) _{\mathrm{S}}} 
      \Bigg\{
      x^\prime _{\mathrm{fp}}
      - (x^\prime \vert x) _{\mathrm{tot}} x _{\mathrm{obj}}
      - (x^\prime \vert x^\prime) _{\mathrm{tot}} x^\prime _{\mathrm{obj}} \\
      - \frac{(x^\prime \vert \delta) _{\mathrm{S}}}{(x \vert \delta) _{\mathrm{S}}}
        [ x _{\mathrm{fp}} - (x \vert x) _{\mathrm{tot}} x _{\mathrm{obj}} ] 
      \Bigg\},
\end{multline*}
where 
\begin{equation*}
  \begin{aligned}
    (x \vert x) _{\mathrm{tot}}
    &= (x \vert x) _{\mathrm{S}} (x \vert x) _{\mathrm{B}} , \\
    (x^\prime \vert x^\prime) _{\mathrm{tot}}
    &= (x^\prime \vert x^\prime) _{\mathrm{S}} (x^\prime \vert x^\prime) _{\mathrm{B}} , \\
    (x^\prime \vert x) _{\mathrm{tot}}
    &= (x^\prime \vert x) _{\mathrm{S}} (x \vert x) _{\mathrm{B}} + (x^\prime \vert x^\prime) _{\mathrm{S}} (x^\prime \vert x) _{\mathrm{B}}, \\
  \end{aligned}
\end{equation*}
with $\mathit{M} _{\mathrm{tot}} \coloneqq \mathit{M} _{\mathrm{S}} \mathit{M} _{\mathrm{B}}$ being the transfer map of the total system consisting of the beamline and the spectrometer.
The content of $\{ \, \}$ of the RHS can be understood as the angle measured at the focal plane with the contributions  from the position, angle, and momentum subtracted from it.

\begin{figure}[t]
  \centering
  \includegraphics{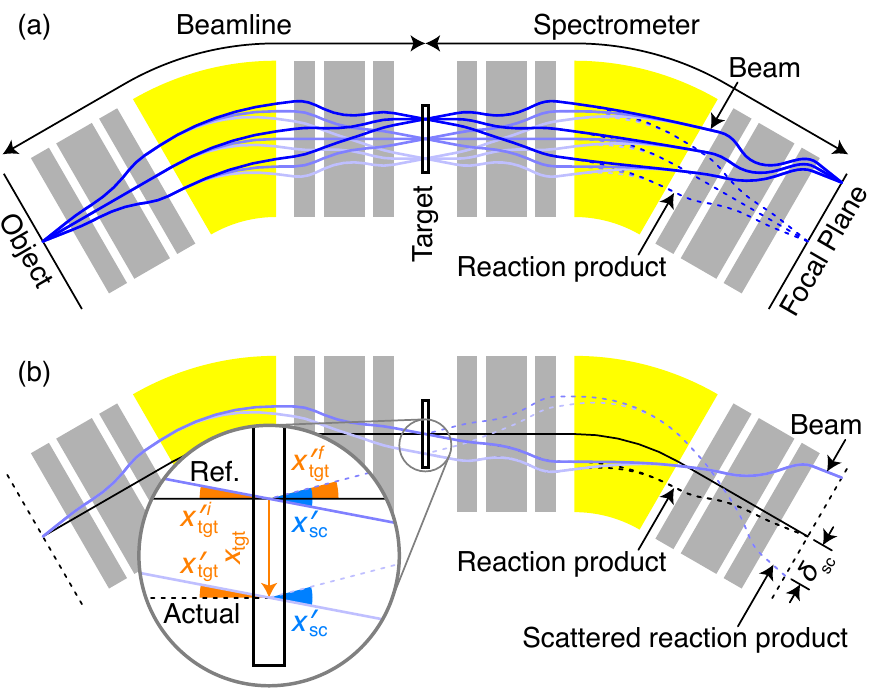}
  \caption{%
  (a) An example of an ion-optical system consisting of a beamline and a spectrometer, each of which is symbolically represented by a dipole magnet sandwiched by triplet quadrupoles on each end.
  The beamline is dispersion-matched to the spectrometer and the entire system is achromatic.
  Three different momentum components are denoted by different colors, each of which is emitted at three different angles.
  The trajectories of the beam particles are denoted by the solid curves, while those of the reaction product, with changes in momentum ($\delta_{\mathrm{sc}}$), are denoted by the thin dotted curves.
  The spectrometer's settings are such that the reaction product is centered in the focal plane.
  The trajectories of the beam come into a dispersive focus at the target position, and then into an achromatic focus at the focal plane.
  The trajectories of the reaction products come into an achromatic focus at the focal plane, with lateral displacement due to the difference in momentum.
  (b) This illustrates scattering of the reaction product which induces changes in the angle ($x^\prime _{\mathrm{sc}}$) and momentum ($\delta _{\mathrm{sc}}$).
  At the focal plane, the scattered reaction product is further displaced laterally due to the momentum change as seen in Eq.~\eqref{eq:dm_xS2_aS2_x}.
  The angle change is obtained as the difference between the reconstructed outgoing angle $x^\prime {}_{\mathrm{tgt}} ^f$ and the incoming angle $x^\prime {}_{\mathrm{tgt}} ^i$, both of which are corrected for the momentum of the beam as seen in Eqs.~\eqref{eq:dm_asc_S0_all}.
  }
  \label{fig:dispersion_matching_analysis}
\end{figure}

By using $\vec{X} _{\mathrm{tgt}} = \mathit{M} _{\mathrm{B}} \vec{X} _{\mathrm{obj}}$, i.e.
\begin{equation*}
  \begin{pmatrix}
    x _{\mathrm{tgt}} \\
    x^\prime _{\mathrm{tgt}} \\
    \delta _{\mathrm{tgt}}
  \end{pmatrix}
  =
  \begin{pmatrix}
  (x \vert x) _{\mathrm{B}} & 0 & (x \vert \delta) _{\mathrm{B}} \\
  (x^\prime \vert x) _{\mathrm{B}} & (x^\prime \vert x^\prime) _{\mathrm{B}} & (x^\prime \vert \delta) _{\mathrm{B}} \\
  0 & 0 & 1
  \end{pmatrix}
  \begin{pmatrix}
  x _{\mathrm{obj}} \\
  x^\prime _{\mathrm{obj}} \\
  \delta _{\mathrm{obj}}
  \end{pmatrix},
\end{equation*}
Eq.~\eqref{eq:dm_X_S2} can be rewritten as
\begin{align*}
  \vec{X}_{\mathrm{fp}} 
  &= \mathit{M} _{\mathrm{S}} (\vec{X}_{\mathrm{tgt}} +
  \vec{X}_{\mathrm{sc}}) \\
  &= 
  \begin{pmatrix}
    (x \vert x) _{\mathrm{S}} & 0 & (x \vert \delta) _{\mathrm{S}} \\
    (x^\prime \vert x) _{\mathrm{S}} & (x^\prime \vert x^\prime) _{\mathrm{S}} & (x^\prime \vert \delta) _{\mathrm{S}} \\
    0 & 0 & 1
  \end{pmatrix}
  \left[
  \begin{pmatrix}
    x _{\mathrm{tgt}} \\
    x^\prime _{\mathrm{tgt}} \\
    \delta _{\mathrm{obj}}
  \end{pmatrix}
  +
  \begin{pmatrix}
    0 \\
    x^\prime _{\mathrm{sc}} \\
    \delta _{\mathrm{sc}}
  \end{pmatrix}
  \right] ,
\end{align*}
which gives
\begin{subequations} \label{eq:dm_xS2_aS2_S0}
\begin{align}
  x _{\mathrm{fp}} \label{eq:dm_xS2_S0}
  ={}& \, (x \vert x) _{\mathrm{S}} x _{\mathrm{tgt}}
      + (x \vert \delta) _{\mathrm{S}} (\delta _{\mathrm{obj}} + \delta _{\mathrm{sc}}) , \\
  x^\prime _{\mathrm{fp}}
  ={}& \, (x^\prime \vert x) _{\mathrm{S}} x _{\mathrm{tgt}}
      + (x^\prime \vert x^\prime) _{\mathrm{S}} (x^\prime _{\mathrm{tgt}} + x^\prime _{\mathrm{sc}})
      + (x^\prime \vert \delta) _{\mathrm{S}} (\delta _{\mathrm{obj}} + \delta _{\mathrm{sc}}) .
\end{align}
\end{subequations}
By eliminating $\delta _{\mathrm{obj}} + \delta _{\mathrm{sc}}$, Eqs.~\eqref{eq:dm_xS2_aS2_S0} can be reduced to,
\begin{align*}
  & (x^\prime \vert x^\prime) _{\mathrm{S}} (x^\prime _{\mathrm{tgt}} + x^\prime _{\mathrm{sc}}) \\
  ={}& x^\prime _{\mathrm{fp}} - \frac{(x^\prime \vert \delta)_{\mathrm{S}}}{(x \vert \delta)_{\mathrm{S}}} x _{\mathrm{fp}}
    + \left[
        \frac{(x^\prime \vert \delta)_{\mathrm{S}} (x \vert x)_{\mathrm{S}}}{(x \vert \delta)_{\mathrm{S}}} - (x^\prime \vert x)_{\mathrm{S}} 
      \right] x _{\mathrm{tgt}} \\
  ={}& x^\prime _{\mathrm{fp}} - \frac{(x^\prime \vert \delta)_{\mathrm{S}}}{(x \vert \delta)_{\mathrm{S}}} x _{\mathrm{fp}}
    - \left[
        \frac{(x^\prime \vert \delta)_{\mathrm{S}}}{(x \vert \delta)_{\mathrm{B}}} + (x^\prime \vert x)_{\mathrm{S}} 
      \right] x _{\mathrm{tgt}} \\
  ={}& x^\prime _{\mathrm{fp}} - \frac{(x^\prime \vert \delta)_{\mathrm{S}}}{(x \vert \delta)_{\mathrm{S}}} x _{\mathrm{fp}}
    + \frac{(x^\prime \vert x^\prime)_{\mathrm{S}} (x^\prime \vert \delta)_{\mathrm{B}}}{(x \vert \delta)_{\mathrm{B}}} x _{\mathrm{tgt}} ,
\end{align*}
where the dispersion-matching conditions \eqref{eq:dm_condition} were used.
$x^\prime _{\mathrm{sc}}$ can be written as a difference
\begin{subequations} \label{eq:dm_asc_S0_all}
\begin{equation} \label{eq:dm_asc_S0}
  x^\prime _{\mathrm{sc}} 
  = x^\prime {}_{\mathrm{tgt}} ^f - x^\prime {}_{\mathrm{tgt}} ^i ,
\end{equation}
where $x^\prime {}_{\mathrm{tgt}} ^f$ and $x^\prime {}_{\mathrm{tgt}} ^i$ are
\begin{align}\label{eq:asc_i_f}
  x^\prime {}_{\mathrm{tgt}} ^f &\coloneqq \,
    \frac{1}{(x^\prime \vert x^\prime) _{\mathrm{S}}} 
    \left[
      x^\prime _{\mathrm{fp}}
      - \frac{(x^\prime \vert \delta)_{\mathrm{S}}}{(x \vert \delta)_{\mathrm{S}}} x _{\mathrm{fp}} 
    \right] , \\
  x^\prime {}_{\mathrm{tgt}} ^i &\coloneqq \,
    x^\prime _{\mathrm{tgt}}
    - \frac{(x^\prime \vert \delta)_{\mathrm{B}}}{(x \vert \delta)_{\mathrm{B}}} x
    _{\mathrm{tgt}} , 
\end{align}
\end{subequations}
This way the scattering angle $x^\prime _{\mathrm{sc}}$ is understood as the difference between the corrected final (outgoing) angle $x^\prime {}_{\mathrm{tgt}} ^f$ and the corrected initial (incoming) angle $x^\prime {}_{\mathrm{tgt}} ^i$.
This is illustrated in Fig.~\ref{fig:dispersion_matching_analysis}.

The angle $x^\prime _{\mathrm{tgt}}$ is the incoming angle at the target position $x _{\mathrm{tgt}}$, which can be measured by the detectors. 
The corrected incoming angle $x^\prime {}_{\mathrm{tgt}} ^i$ is with the correction for the angular dispersion at the target position, $x _{\mathrm{tgt}} \cdot (x^\prime \vert \delta)_{\mathrm{B}} / (x \vert \delta)_{\mathrm{B}}$, which is due to from the momentum deviation in the beamline.

The corrected outgoing angle $x^\prime {}_{\mathrm{tgt}} ^f$ is the angle at the focal plane, $x^\prime _{\mathrm{fp}}$, corrected for the angular dispersion due to the momentum deviation in the spectrometer, $x _{\mathrm{fp}} \cdot (x^\prime \vert \delta)_{\mathrm{S}} / (x \vert \delta)_{\mathrm{S}}$, which is then divided by the angular magnification, $(x^\prime \vert x^\prime) _{\mathrm{S}}$.
$x^\prime {}_{\mathrm{tgt}} ^f$ is the angle after the target reconstructed in first order using the position and angle obtained at the focal plane together with the ion-optical properties (the elements of the transfer maps) of the spectrometer, where the particle is assumed to be emitted from $x _{\mathrm{tgt}} = 0$.

In both $x^\prime {}_{\mathrm{tgt}} ^f$ and $x^\prime {}_{\mathrm{tgt}} ^i$, there are also contributions from $x _{\mathrm{obj}}$, which need to be subtracted from them, namely,
\begin{equation*}
  \begin{aligned}
  \tilde{x^\prime} _{\mathrm{tgt}} ^f &\coloneqq \,
    \frac{1}{(x^\prime \vert x^\prime) _{\mathrm{S}}} 
    \Bigg[ 
      \left\{ x^\prime _{\mathrm{fp}} - (x^\prime \vert x) _{\mathrm{tot}} x _{\mathrm{obj}} \right\} \\
      - \frac{(x^\prime \vert \delta)_{\mathrm{S}}}{(x \vert \delta)_{\mathrm{S}}} 
      \left\{ x _{\mathrm{fp}} - (x \vert x)_{\mathrm{tot}} x _{\mathrm{obj}} \right\}
    \Bigg] , \span \omit \\
  \tilde{x^\prime} _{\mathrm{tgt}} ^i &\coloneqq \,
    \left\{ x^\prime _{\mathrm{tgt}} - (x^\prime \vert x) _{\mathrm{B}} x _{\mathrm{obj}} \right\}
    - \frac{(x^\prime \vert \delta)_{\mathrm{B}}}{(x \vert \delta)_{\mathrm{B}}} 
    \left\{ x _{\mathrm{tgt}} - (x \vert x)_{\mathrm{B}} x _{\mathrm{obj}}\right\}.
  \end{aligned}
\end{equation*}
However, $x^\prime {}_{\mathrm{tgt}} ^f$ and $x^\prime {}_{\mathrm{tgt}} ^i$ instead of $\tilde{x^\prime} _{\mathrm{tgt}} ^f$ and $\tilde{x^\prime} _{\mathrm{tgt}} ^i$ can be used to obtain the correct value of $x^\prime _{\mathrm{sc}}$ in first order because the $x _{\mathrm{obj}}$ terms cancel out in the final expression of $x^\prime _{\mathrm{sc}}$.
Neither $x^\prime {}_{\mathrm{tgt}} ^i$ nor $x^\prime {}_{\mathrm{tgt}} ^f$ contains the bare $x _{\mathrm{obj}}$ or $x^\prime _{\mathrm{obj}}$, but in the form of $x _{\mathrm{tgt}}$ and $x^\prime _{\mathrm{tgt}}$.
This indicates that the position and angle measurements at the target location alone is sufficient to deduce the scattering angle, and that at the object is not necessary.

\subsection{Partial dispersion matching}

Let us now consider the case where the beam is dispersion-matched to an intermediate focal plane instead of the final focal plane.
Let the position, angle, and momentum at the intermediate focal plane (F1) and the final focal plane (F2) be $\vec{X}_1 = (x_1,x^\prime_1,\delta_1)$ and $\vec{X}_2 = (x_2,x^\prime_2,\delta_2)$, respectively ($\delta _1 = \delta _2 = \delta _{\mathrm{obj}} + \delta _{\mathrm{sc}}$).
Let the transfer map from the target (F0) to F1 and that from F1 to F2 be $\mathit{M} _{10}$ and $\mathit{M} _{21}$, respectively, such that
\begin{align*}
  \vec{X}_1 &= \mathit{M} _{10} \vec{X} _{\mathrm{tgt}}, \\
  \vec{X}_2 &= \mathit{M} _{21} \vec{X} _1 
            = \mathit{M} _{21} \mathit{M} _{10} \vec{X} _{\mathrm{tgt}}
            = \mathit{M} _{\mathrm{S}} \vec{X} _{\mathrm{tgt}}.
\end{align*}
Because the position at the target $x _{\mathrm{tgt}}$ contains information about the momentum deviation in the beam $\delta _{\mathrm{obj}}$ as
\begin{equation*}
  x _{\mathrm{tgt}} = (x \vert x) _{\mathrm{B}} x _{\mathrm{obj}} + (x \vert \delta) _{\mathrm{B}} \delta _{\mathrm{obj}},
\end{equation*}
$\delta _{\mathrm{obj}}$ can be obtained from $x _{\mathrm{tgt}}$ as
\begin{equation*}
  \delta _{\mathrm{obj}}
  = \frac{x _{\mathrm{tgt}}}{(x \vert \delta) _{\mathrm{B}}} - \frac{(x \vert x) _{\mathrm{B}}}{(x \vert \delta) _{\mathrm{B}}} x _{\mathrm{obj}}.
\end{equation*}
Eq.~\eqref{eq:dm_xS2_S0} is now rewritten as
\begin{equation*}
  x _2 
  = \, (x \vert x) _{\mathrm{S}} x _{\mathrm{tgt}}
      + (x \vert \delta) _{\mathrm{S}} (\delta _{\mathrm{obj}} + \delta _{\mathrm{sc}}) ,
\end{equation*}
and the momentum change $\delta _{\mathrm{sc}}$ due to the reaction is
\begin{equation*}
 \delta _{\mathrm{sc}}
  = \left[ \frac{x _2}{(x \vert \delta)_{\mathrm{S}}} + \frac{(x \vert x)_{\mathrm{B}}}{(x \vert \delta)_{\mathrm{B}}} x _{\mathrm{obj}} \right]
  - \left[ \frac{1}{(x \vert \delta) _{\mathrm{B}}} + \frac{(x \vert x)_{\mathrm{S}}}{(x \vert \delta)_{\mathrm{S}}} \right] x _{\mathrm{tgt}}.
\end{equation*}
The second $x _{\mathrm{tgt}}$ term on the RHS has been added to Eq.~\eqref{eq:dm_deltasc}, due to the fact that the beamline is not dispersion-matched to F2 but only to F1.

The scattering angle is obtained similarly to Eq.~\eqref{eq:dm_asc_S0} as
\begin{equation*}
  x^\prime _{\mathrm{sc}}
  = x^\prime {}_{\mathrm{tgt}} ^f - x^\prime {}_{\mathrm{tgt}} ^i
\end{equation*}
with
\begin{align*}
  x^\prime {}_{\mathrm{tgt}} ^i &\coloneqq \,
    x^\prime _{\mathrm{tgt}}
    - \frac{(x^\prime \vert \delta)_{\mathrm{B}}}{(x \vert \delta)_{\mathrm{B}}} x
    _{\mathrm{tgt}} , \\
  x^\prime {}_{\mathrm{tgt}} ^f &\coloneqq \,
    \frac{1}{(x^\prime \vert x^\prime) _{10}} 
    \left[
      x^\prime _1
      - \frac{(x^\prime \vert \delta)_{10}}{(x \vert \delta)_{10}} x _1
    \right] .
\end{align*}
Here, $x^\prime {}_{\mathrm{tgt}} ^i$ is identical to that in Eq.~\eqref{eq:asc_i_f}, but $(x _{\mathrm{fp}},x^\prime _{\mathrm{fp}})$ in $x^\prime {}_{\mathrm{tgt}} ^f$ are replaced by $(x _1,x^\prime _1)$, and $(x^\prime \vert x^\prime)_{\mathrm{S}}$ by $(x^\prime \vert x^\prime)_{10}$.

With
\begin{align*}
  (x \vert \delta) _{\mathrm{S}} ={}& (x \vert x) _{21} (x \vert \delta) _{10} + (x \vert \delta) _{21}, \\
  x _2
  ={}& \, (x \vert x) _{21} x _1
    + (x \vert \delta) _{21} (\delta _{\mathrm{obj}} + \delta _{\mathrm{sc}}) , \\
  x^\prime _2
  ={}& \, (x^\prime \vert x) _{21} x _1
    + (x^\prime \vert x^\prime) _{21} x^\prime _1
    + (x^\prime \vert \delta) _{21} (\delta _{\mathrm{obj}} + \delta _{\mathrm{sc}}) , \\
  \delta _{\mathrm{sc}}
  ={}& \, \frac{x _1}{(x \vert \delta)_{10}} + \frac{(x \vert x) _{\mathrm{B}}}{(x \vert \delta) _{\mathrm{B}}} x _{\mathrm{obj}} ,
\end{align*}
$x^\prime {}_{\mathrm{tgt}} ^f$ can be reduced to
\begin{multline} \label{eq:partialDM_angle}
x^\prime {}_{\mathrm{tgt}} ^f
= \frac{1}{(x^\prime \vert x^\prime) _{\mathrm{S}}} 
  \left\{
    \left[ x^\prime _2 - (x^\prime \vert \delta) _{21} \frac{x _{\mathrm{tgt}}}{(x \vert \delta) _{\mathrm{B}}} 
    \right] \right. \\
    \left. 
    - \frac{(x^\prime \vert \delta)_{\mathrm{S}}}{(x \vert \delta)_{\mathrm{S}}} 
    \left[ x _2 - (x \vert \delta) _{21} \frac{x _{\mathrm{tgt}}}{(x \vert \delta) _{\mathrm{B}}} \right]
  \right\} .
\end{multline}
As in the case of the full dispersion matching, the angle $x^\prime _{\mathrm{sc}}$ can be deduced from the position and angle at the focal plane and those at the target location, but in the present case $x _{\mathrm{tgt}}$ goes into $x^\prime {}_{\mathrm{tgt}} ^f$ in addition to $x^\prime {}_{\mathrm{tgt}} ^i$.
Eq.~\eqref{eq:partialDM_angle} implies that the inverse mapping is applied to 
$x _2 - x _{\mathrm{tgt}} \cdot (x \vert \delta) _{21} / (x \vert \delta) _{\mathrm{B}}$
and 
$x^\prime _2 - x _{\mathrm{tgt}} \cdot (x^\prime \vert \delta) _{21} / (x \vert \delta) _{\mathrm{B}}$,
instead of the bare $x _2$ and $x^\prime _2$, whereby the effects of the momentum deviation of the beam, $\delta _{\mathrm{obj}}$, are eliminated from these variables.

\begin{figure}[t]
  \centering
  \includegraphics{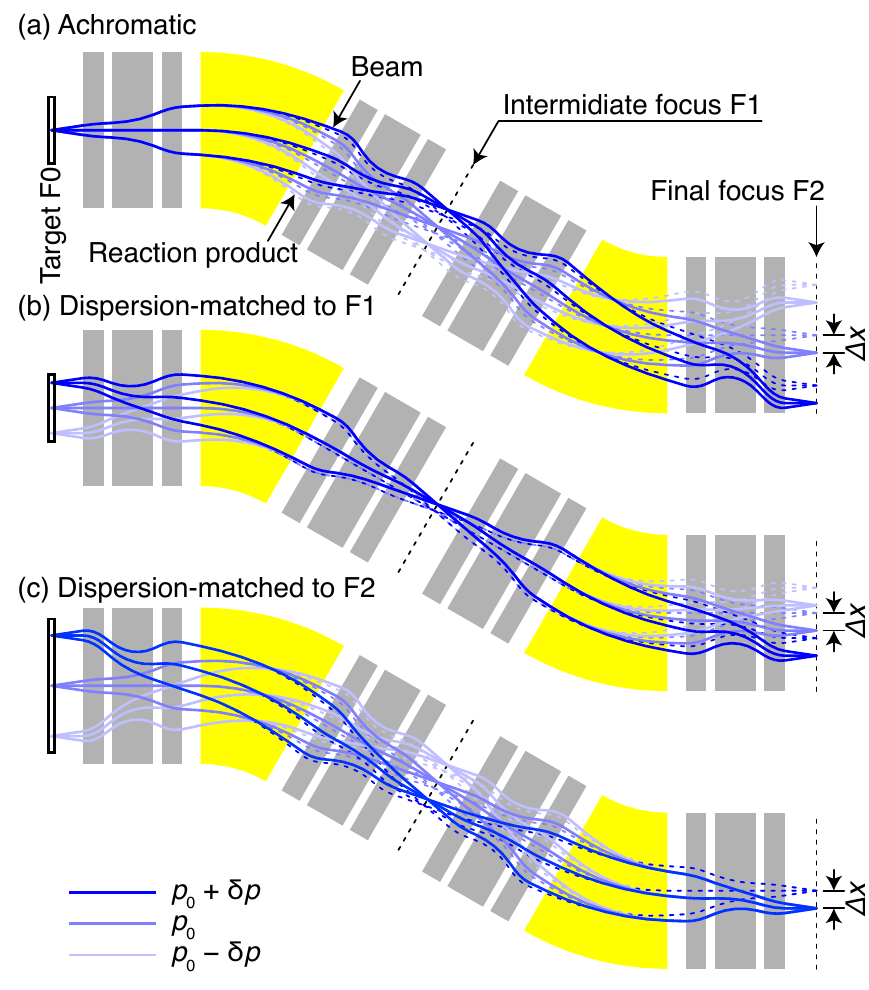}
  \caption{%
  Illustration of the beam transport in the spectrometer comparing the cases where (a) the beam is achromatic at the target, (b) it is dispersion-matched only to F1, and (c) it is dispersion-matched fully to F2.
  Three different momentum components are denoted by different colors, but they are the same for all the three figures.
  For each momentum, trajectories for three angles are depicted.
  The trajectories of the beam particles are denoted by the solid curves, while those of the reaction product, with changes in momentum ($\delta_{\mathrm{sc}}$), are denoted by the thin dotted curves.
  The spectrometer's settings are such that the reaction product is centered in the final focus FS2.
  }
  \label{fig:dispersion_matching_comparison}
\end{figure}

Figure~\ref{fig:dispersion_matching_comparison} illustrates the ion optics in the spectrometer in the achromatic, partially dispersion-matched, fully dispersion-matched cases.

\pdfbookmark{\refname}{References}

\end{document}